\newif\ifOneCol

\OneColtrue
\ifOneCol
	\documentclass[12pt, draftclsnofoot, onecolumn]{IEEEtran}
\else
	\documentclass[10pt, doublecolumn]{IEEEtran}
\fi 
\ifCLASSINFOpdf
\else
\fi
\usepackage{subfloat}
\usepackage{adjustbox}
\usepackage{enumerate}
\usepackage{amsmath}
\usepackage{amsthm}
\usepackage{amsfonts,amssymb}
\usepackage{color}
\usepackage{url}
\usepackage{tikz}
\usepackage[english]{babel}
\usepackage[T1]{fontenc}
\usepackage[ansinew]{inputenc}
\usepackage{graphicx}
\usepackage{algorithm}
\usepackage{algpseudocode}
\usepackage{pgfplots}
\usepackage{enumitem}
\usepackage{caption}
\usepackage[normalem]{ulem}
\usepackage{varwidth}
\usepackage[noadjust]{cite}
\usepackage[labelsep=quad,indention=10pt]{subfig}
\usepackage[outdir=./]{epstopdf}
\usepackage{xcolor}
\usepackage{hyperref}
\usepackage{upquote}
\usepackage{textcomp}
\usepackage[normalem]{ulem}
\usepackage{comment}

\newcommand{\fGr}[1]{\boldsymbol{#1}}
\newcommand{\fat}[1]{\boldsymbol{#1}}

\newcommand{\KL}[2]{\mathrm{D}\left[#1\left\lVert\right. #2\right]}
\newcommand{\bi}[2]{b_{\mathrm{#1},{#2}}}

\allowdisplaybreaks
\pgfplotscreateplotcyclelist{mylist1}{
	{blue,mark=*},	
	{teal,mark=star},
	{green!60!black,mark=o},
	{cyan,mark=+,mark size=2.5},
	{red,mark=square},
	{brown!60!black,mark options={fill=brown!40},mark=otimes*}}
\pgfplotscreateplotcyclelist{mylist}{
	{blue,mark=*},
	{brown!60!black,mark options={fill=brown!40},mark=otimes*},
	{green!60!black,mark=o},
	{red,mark=square},
	{cyan,mark=+,mark size=2.5}}

 \usepackage{setspace}
 \let\Algorithm\algorithm
 \renewcommand\algorithm[1][]{\Algorithm[#1]\setstretch{1.3}}
 
 \definecolor{mycolor}{rgb}{0, 0, 0}
 
 \ifOneCol
\else
\fi 

\usetikzlibrary{shapes.geometric}
\usetikzlibrary{intersections}
\pgfplotsset{compat=1.3}
\makeatletter
\def\parsenode[#1]#2\pgf@nil{%
	\tikzset{label node/.style={#1}}
	\def\nodetext{#2}
}

\tikzset{
	add node at x/.style 2 args={
		name path global=plot line,
		/pgfplots/execute at end plot visualization/.append={
			\begingroup
			\@ifnextchar[{\parsenode}{\parsenode[]}#2\pgf@nil
			\path [name path global = position line #1-1]
			({axis cs:#1,0}|-{rel axis cs:0,0}) --
			({axis cs:#1,0}|-{rel axis cs:0,1});
			\path [xshift=1pt, name path global = position line #1-2]
			({axis cs:#1,0}|-{rel axis cs:0,0}) --
			({axis cs:#1,0}|-{rel axis cs:0,1});
			\path [
			name intersections={
				of={plot line and position line #1-1},
				name=left intersection
			},
			name intersections={
				of={plot line and position line #1-2},
				name=right intersection
			},
			label node/.append style={pos=1}
			] (left intersection-1) -- (right intersection-1)
			node [label node]{\nodetext};
			\endgroup
		}
	}
}
\makeatletter
\def\BState{\State\hskip-\ALG@thistlm}
\makeatother
\pgfplotsset{compat=1.5}
\begin{document}
	\title{Unifying Message Passing Algorithms\\ \vspace{-0.3cm} Under the Framework of \\\vspace{-0.3cm}Constrained Bethe Free Energy Minimization}

	\author{\large
		\IEEEauthorblockN{
			\normalsize{
				Dan~Zhang, 
				Xiaohang~Song, \textit{Member, IEEE}, 
				Wenjin Wang, \textit{Member, IEEE}
				\\Gerhard~Fettweis, \textit{Fellow, IEEE}, 
				and Xiqi Gao, \textit{Fellow, IEEE}}}
				\vspace*{-10mm}
				\thanks{This work has been supported jointly by the German Science Foundation (DFG) and the National Natural Science Foundation of China (NSFC) under the project "Large-Scale and Hierarchical Bayesian Inference for Future Mobile Communication Networks".}
\thanks{D.~Zhang was with the Vodafone Chair Mobile Communications Systems, Technische Universit\"at Dresden, Germany, and is currently with Bosch Center for Artificial Intelligence, Germany, Email: dan.zhang2@de.bosch.com. The work was conducted at her stay with the Vodafone Chair.}
\thanks{X.~Song and G.~Fettweis are with the Vodafone Chair Mobile Communications Systems, Technische Universit\"at Dresden, Germany, Email: \{xiaohang.song, gerhard.fettweis\}@tu-dresden.de.}
\thanks{W.~Wang and X.~Gao are with National Mobile Communications Research Laboratory, Southeast University, China, Email: \{wangwj, xqgao\}@seu.edu.cn.}
\thanks{We thank Dr. Li You from Southeast University, China and Prof. Bernard H. Fleury from Aalborg University, Denmark, for the insightful discussion that greatly assisted this work.}
\thanks{Code of this work can be found at: \url{https://github.com/XiaohangSong/EPV-SSR}.}
}
	\markboth{}{}%

	\maketitle
	\vspace{-0.9cm}
	\begin{abstract}
\vspace{-0.3cm}Variational message passing (VMP), belief propagation (BP) and expectation propagation (EP) have found their wide applications in complex statistical signal processing problems. In addition to viewing them as a class of algorithms operating on graphical models, this paper unifies them under an optimization framework, namely, Bethe free energy minimization with differently and appropriately imposed constraints. This new perspective in terms of constraint manipulation can offer additional insights on the connection between different message passing algorithms and is valid for a generic statistical model. It also founds a theoretical framework to systematically derive message passing variants. Taking the sparse signal recovery (SSR) problem as an example, a low-complexity EP variant can be obtained by simple constraint reformulation, delivering better estimation performance with lower complexity than the standard EP algorithm. Furthermore, we can resort to the framework for the systematic derivation of hybrid message passing for complex inference tasks. Notably, a hybrid message passing algorithm is exemplarily derived for joint SSR and statistical model learning with near-optimal inference performance and scalable complexity. 
	\end{abstract}	
\vspace{-0.3cm}
	\begin{IEEEkeywords}
		\vspace{-0.3cm}Statistical inference, Bethe free energy, message passing algorithms, constrained optimization. \vspace{-0.3cm}
	\end{IEEEkeywords}

	%
	\IEEEpeerreviewmaketitle

 \ifOneCol
	\newpage
\else
\fi 	
\section{Introduction}

Advances in signal processing have always been an enabler of innovations in the evolution of mobile communications networks. Many signal processing problems can be formulated as a statistical inference task of estimating the latent random variable $\fat{x}$ given the realization of a statistically related observation $\fat{y}$. Knowing the likelihood function $p(\fat{y}\vert\fat{x})$ and the prior density $p(\fat{x})$, the inference task, performing under the Bayesian framework, relies on the a-posteriori density $p(\fat{x}\vert\fat{y})\propto p(\fat{y}\vert\fat{x})p(\fat{x})$. Treating $\fat{x}$ and $\fat{y}$ as the input and output of a system, respectively, the a-posteriori density mathematically describes the input-output statistical dependence, permitting inference under various criteria, e.g., minimum mean square error (MMSE) $\fat{\hat{x}}=\mathrm{E}[\fat{x}\vert \fat{y}]$, or maximum a-posteriori (MAP) $\fat{\hat{x}}=\arg\max_{\fat{x}}p(\fat{x}\vert\fat{y})$. 

In practical large-scale wireless communications systems, it might be over complicated to evaluate $p(\fat{x}\vert\fat{y})$ or to derive its statistical properties such as MMSE and MAP estimates. The reason could be a too large feasible space of $\fat{x}$, or the form of $p(\fat{x}\vert\fat{y})$ is analytically intractable. To cope with such cases, one can resort to some form of approximations that generally fall into two classes, i.e., deterministic and stochastic approximations. As mentioned in~\cite{Bishop2006,Barber2012}, stochastic approximation approaches, e.g., the Markov Chain Monte Carlo, tend to be more computationally demanding.\footnote{Here we note that no single approximation technique, neither deterministic nor stochastic, outperforms all others on all problems. In fact, both types of approximations are broad enough research topics to be studied on their own.} Aiming at low complexity methods for large-scale systems, in this work, we consider a family of deterministic approximations termed variational Bayesian inference.\looseness=-1

\subsection{Variational Bayesian inference and message passing}

Briefly, variational Bayesian inference attempts to approximate $p(\fat{x}\vert\fat{y})$ by an alternative density $\hat{b}(\fat{x})$. By constraining the form of $\hat{b}(\fat{x})$, one can ensure the mathematical tractability of deriving statistical properties on top of it. In the meantime, sufficient approximation accuracy must be ensured to perform inference. To this end, the Kullback-Leibler (KL) divergence (a.k.a. relative entropy), which quantifies the difference between a given density pair, is used here. Limiting to a family $\mathcal{Q}$ of densities in the desired form, $\hat{b}(\fat{x})$ is chosen to yield the minimal KL divergence with respect to $p(\fat{x}\vert\fat{y})$~\cite{Jordan1999}. In the context of physics, such an optimization problem is also known as variational free energy minimization (a.k.a. Gibbs free energy minimization)~\cite{Yedidia2005}.

There are two well-accepted approaches to construct the specialized density family $\mathcal{Q}$. Firstly, the mean field approach defines it as a set of fully factorisable densities, and variational message passing (VMP)~\cite{Winn05} (a.k.a. mean field algorithm) was proposed accordingly as an iterative solution. Here, we note that the expectation maximization (EM) algorithm initially introduced to solve the maximum likelihood (ML) estimation problem~\cite{Dempster77} can be considered as a special case of VMP. It additionally forces the densities to be Dirac-delta functions with a single parameter. The second approach constructs $\mathcal{Q}$ by exploiting the factorization of $p(\fat{x}\vert \fat{y})$ or $p(\fat{x},\fat{y})$.\footnote{Here, the density $p(\fat{y})$ involved in $p(\fat{x},\fat{y})=p(\fat{x}\vert\fat{y})p(\fat{y})$ is treated as a constant as it is not a function of the variable $\fat{x}$.} The variational free energy is then approximated by the so-called Bethe free energy~\cite{Yedidia2005}, which can be minimized by belief propagation (BP)~\cite{Pearl}. However, BP is not well suited to accomplish tasks that involve continuous random variables\footnote{\textcolor{mycolor}{Exception would be cases where the beliefs follow distributions that are closed under marginalization and multiplication, such as linear Gaussian models where the beliefs of BP over iterations can remain Gaussian.}}, e.g., synchronization and channel estimation in communications systems. To tackle this issue, expectation propagation (EP) adds one step to BP, i.e., projecting the beliefs onto a specific function family with the corresponding sufficient statistics for analytical tractability~\cite{Minka2001}. Being the approximate solutions to variational free energy minimization, the above mentioned algorithms have been widely applied to solve problems that arise in communication systems with large dimensions, e.g., multiuser detection~\cite{LLiu2019, LLiu2019_b}, coded modulation capacity analysis~\cite{Zhang2016SPL}, MIMO channel estimation and data detection~\cite{Hansen_Fleury_2018, Heath_2019}.

Approximate message passing (AMP) was developed in~\cite{Donoho10112009} for compressed sensing and later generalized by~\cite{Rangan2011} (thereby termed GAMP) for solving the problem in generalized linear systems. They exhibit intrinsic connections to EP in the large system limit, e.g., circumventing matrix inversion in EP with the aid of the self-averaging method~\cite{Cakmak2014} or neglecting high-order terms when computing EP messages~\cite{Meng2015, new_insight_GAMP}.\footnote{The EP algorithms in work~\cite{Cakmak2014} and~\cite{Meng2015} are w.r.t. two different types of factorization on the same objective function.} Due to the relatively low complexity, they have become pragmatic alternatives to BP and EP in large systems, e.g.,~\cite{sWu2014,sWang2015a, Lyu_2019}. Under i.i.d. Gaussian MIMO channels, the optimality of (G)AMP for large-scale MIMO detection was assessed in~\cite{Jeon2015} and \cite{capacity_AMP_CM}. Besides, the (G)AMP algorithms were extended to allow more general classes of channels~\cite{VAMP_2017, GVAMP_2016, OAMP_2017, GrAMP_2018, FanHao2017SEU}, e.g., rank-deficient and/or coefficient-correlation channels.

Large-scale sparse signal recovery (SSR) is relevant to communication systems that are under-determined, e.g., active user detection in massive machine-type communications (mMTC) and channel estimation in millimeter wave communications. Among the existing techniques, one class based on the empirical Bayesian framework is termed sparse Bayesian learning (SBL). Both VMP (including its special case EM) and (G)AMP are applicable~\cite{Shutin2011,Schniter2013,Rangan2012_AMP_EM, GGAMP_SBL_2018}. Therefore, they have been applied in the literature for activity detection on random access channels~\cite{Utkovski2017, mutual_broadcast}, estimating sparse channels~\cite{Mo2014, Karseras2015, Stockle2016}, spatial modulation~\cite{Wu_EM_AMP_16}, and joint user activity detection and channel estimation~\cite{Joint_channel_user_activity_one, Joint_channel_user_activity_two}. {\color{mycolor} The convergence of (G)AMP is case-dependent. It can easily diverge when the channel is rank-deficient and/or correlated. Therefore, the work \cite{FanHao2017SEU,Luo_UTAMP} introduced a linear preprocessing step to de-correlate the observations and reduce the system dimension.}\looseness=-1

\subsection{Motivation and contribution of this work}
From the above state-of-the-art overview, we notice that most existing theoretical works investigated message passing algorithms individually, even though their heuristic combinations have already found its applications, e.g.,~\cite{Senst2011a,Badiu2012,Sun2015,wWang2016,NWU2017}. With a joint use of the mean field and Bethe approaches for approximating the variational free energy, VMP and BP were merged in~\cite{Riegler2013} via partitioning the factors into two corresponding subsets. Apart from that, to our best knowledge, little results have been reported on unifying message passing algorithms {\color{mycolor} from an optimization perspective. This paper aims at an optimization framework that can link different message passing algorithms under a generic statistical model}. The main contributions are summarized as follows: 
\begin{itemize}
\item We construct the framework based on constrained Bethe free energy minimization. In particular, the Bethe free energy, an approximation to the variational free energy, is used as the objective function. On top of it, we introduce a set of constraint formulation methods such that BP, EP, and VMP can be analytically attributed to corresponding constrained Bethe free energy minimization. 
\item From this novel perspective of constraint manipulation, we systematically derive new message passing variants, in particular hybrid ones for complex statistical inference problems. It is noted that under our framework, BP and VMP can be combined in a more generalized manner than that in~\cite{Riegler2013}. To further ease the understanding and implementation of hybrid message passing, the conventional factor graph is adapted accordingly for visualization.
\item We exemplarily address a classic SSR problem under the developed framework. Through constraint reformulation, we derive new message passing variants with and without perfect knowledge of the statistical model, outperforming multiple benchmark algorithms in both performance and complexity. 
\item We briefly showcase our framework for solving practical problems in advanced communications scenarios such as mMTC, massive MIMO and millimeter-wave communications.
\end{itemize}
Here, we note that given the space limit the primary focus of the paper is on presenting a generic framework that is applicable to solve inference problems in modern wireless communications systems. Therefore, we choose a simple yet representative application to showcase the application of our framework. For readers that are interested in practical wireless communications problems, we refer to our parallel works~\cite{Wang2017SEU,FanHao2017SEU}.

This paper is organized as follows: Section~\ref{sec_vb} introduces the message passing algorithms, i.e., BP, EP and VMP, for variational free energy minimization, respectively. The key part of the paper is given in Section~\ref{sec_framework}. It describes the optimization framework that can unify BP, EP and VMP under different constraints. In Section~\ref{sec_sbl}, a SSR problem is considered as an example for practicing the developed framework, and four wireless applications are also provided with this framework being their (new) viewpoints. Finally, a conclusion and outlook are presented in Section~\ref{Sec_con}.\looseness=-1
\section{Variational Free Energy Minimization}\label{sec_vb}
 In this section, we link the BP, EP and VMP to variational free energy minimization~\cite{Yedidia2005}. The derivation starts with a density $p(\fat{x}\vert\fat{y})$ constructed from a non-negative function $f(\fat{x})$ as
\begin{align}
p(\fat{x}\vert\fat{y})=\frac{1}{Z} f(\fat{x})\quad\text{with}\,\, Z = \int f(\fat{x})\mathrm{d}\fat{x},
\end{align}
where $\mathrm{d} \fat{x}$ denotes the Lebesgue measure and counting measure respectively for continuous and discrete case. Using a trial density $b(\fat{x})$, we define a new function
\begin{align}
	F(b) & \triangleq \int b(\fat{x})\ln b(\fat{x})\mathrm{d}\fat{x} - \int b(\fat{x}) \ln f(\fat{x})\mathrm{d}\fat{x} = -\ln Z + \mathrm{D}\left[b(\fat{x})\Vert p(\fat{x}\vert\fat{y})\right],\label{F}
\end{align}
where $D\left[\cdot \Vert \cdot\right]$ stands for the KL divergence between two distributions. %
In physics, it is known as the variational free energy of the system and $-\ln Z$ is termed Helmholtz free energy. By minimizing $F(b)$ over all possibilities of the trial density $b(\fat{x})$, the solution is straightforward, i.e., $\hat{b}(\fat{x})=p(\fat{x}\vert\fat{y})$ and $F(\hat{b})=-\ln Z$. Therefore, the minimization of $F(b)$ is an exact procedure to compute $-\ln Z$ and recover $p(\fat{x}\vert\fat{y})$. However, it is not always tractable. One common approximate solution is to limit the feasible set of $b(\fat{x})$ to a density family $\mathcal{Q}$, i.e.,
\begin{align}
\hat{b}(\fat{x})=\arg\min_{b\in\mathcal{Q}}F(b).\label{min_var}
\end{align}
Apparently, the choice of $\mathcal{Q}$ determines the fidelity and tractability of the resulting approximation $\hat{b}(\fat{x})\approx p(\fat{x}\vert \fat{y})$ and $F(\hat{b})\approx-\ln Z$. In the following, we will show how to construct $\mathcal{Q}$ such that BP, EP and VMP are iterative solutions to the corresponding minimization problem.

\subsection{Mean field approximation}
The mean field approximation to the variational free energy minimization problem (\ref{min_var}) constructs $\mathcal{Q}$ as a collection of fully factorizable densities, i.e.,
\begin{align}
b(\fat{x})=\prod_i b_i(x_i)\quad \forall \,\, b(\fat{x})\in\mathcal{Q}.\label{F_vmp}
\end{align}
With such a family, the primary problem (\ref{min_var}) becomes
\begin{align}
\{\hat{b}_i(x_i)\}&=\arg\min_{\{b_i(x_i)\}}F\left (\prod_i b_i(x_i)\right). \label{min_mf}
\end{align}
The problem (\ref{min_mf}) is convex with respect to each individual $b_i(x_i)$, but non-convex when jointly considering $\{b_i(x_i)\}$. The optimal update equation for $b_i(x_i)$ with the others being fixed is
\begin{align}
b_i(x_i)\propto  e^{\int \ln f(\fat{x}) \prod_{i'\neq i} b_{i'}(x_{i'})\mathrm{d}\fat{x}_{\backslash i}},\label{vmp_rule}
\end{align}
where the integral is with respect to $\fat{x}$ except $x_i$. With proper initialization, $\{b_i(x_i)\}$ can then be successively and iteratively updated using (\ref{vmp_rule}) until a local minimum or a saddle point is reached. We note that \eqref{vmp_rule} is identical to the message update rule of VMP~\cite{Winn05}, implying VMP as an iterative solution to \eqref{min_mf} that ignores the statistical dependence between variables at inference. 

\ifOneCol
\begin{figure}[t!]
	\centering
	\includegraphics[scale=0.65]{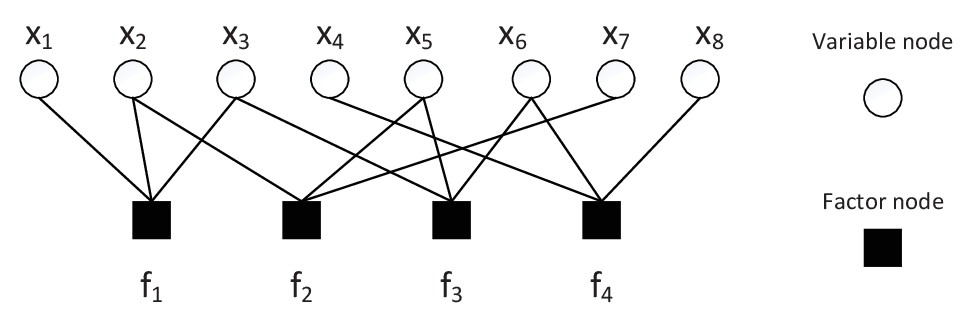}\vspace{-0.3cm}
	\caption{Factor graph of the function $f(\fat{x})=f_1(x_1,x_2,x_3)f_2(x_2,x_5,x_7)f_3(x_3,x_5,x_6) f_4(x_4,x_6,x_8)$.}\vspace{-0.6cm}\label{fig_fg}
\end{figure}
\else
\begin{figure}[t!]
	\includegraphics[scale=0.5]{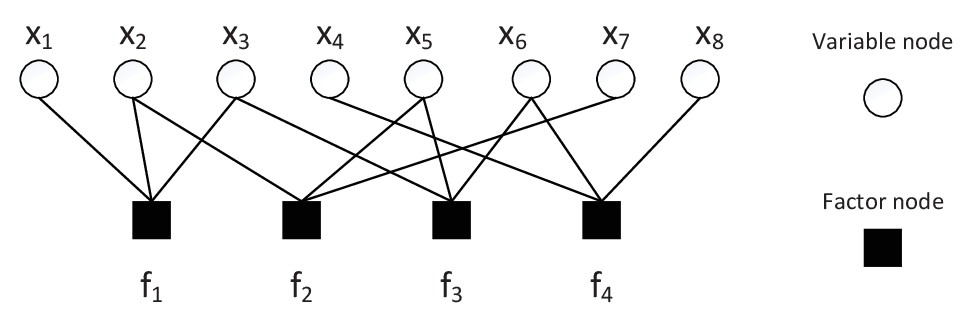}\vspace{-0cm}
	\caption{An illustration of a factor graph that depicts the function $f(\fat{x})=f_1(x_1,x_2,x_3)f_2(x_2,x_5,x_7)f_3(x_3,x_5,x_6) f_4(x_4,x_6,x_8)$.}\vspace{-0.2cm}\label{fig_fg}
\end{figure}
\fi
\subsection{Bethe approximation}
The Bethe approximation attempts to preserve the local dependence during inference by exploiting the intrinsic factorization of $f(\fat{x})$, namely,
\begin{align}
f(\fat{x}) = \prod_a f_a(\fat{x}_a),\label{f_fac}
\end{align}
where the argument $\fat{x}_a$ of the factor function $f_a(\fat{x}_a)$ is a subvector of the vector $\fat{x}$. We can visualize the factorization by means of factor graph~\cite{Kschischang2001}. As the example being depicted in Fig.~\ref{fig_fg}, each entry of $\fat{x}$ is depicted as a variable node, while factor nodes denote the factor functions $\{f_a\}$. One factor node is connected to a set of variable nodes that are its arguments. The variables that are connected to the same factor node are locally dependent. The Bethe approximation introduces the auxiliary densities $\{b_a(\fat{x}_a)\}$ to describe their statistical relations {\color{black} at factor nodes $\{f_a\}$, respectively}. In addition, it defines {\color{black}the marginal beliefs} $\{b_i(x_i)\}$ for each variable node. Aiming at the global dependence across multiple factor nodes, the Bethe approximation requires $\{b_a(\fat{x}_a)\}$ and $\{b_i(x_i)\}$ to fulfill the marginalization consistency constraint
\begin{align}
\int b_a(\fat{x}_a) \mathrm{d} \fat{x}_{a\backslash i}=b_i(x_i)\quad \forall i \,\forall \,a \in \mathcal{A}_i,\label{Bethe_constr}
\end{align}
{\color{mycolor}where the integral is w.r.t. $\fat{x}_a$ except $x_i$, and $\mathcal{A}_i$ collects the indices of the factor functions with $x_i$ being one of their arguments.}

If the factor graph of $f(\fat{x})$ has a tree structure, the optimal setting for $\mathcal{Q}$ is
\begin{align}
b(\fat{x})=\frac{\prod_a b_a(\fat{x}_a)}{\prod_{i} \left[b_i(x_i)\right]^{A_i-1}}\quad\quad \forall\,\, b(\fat{x})\in\mathcal{Q}, \label{Bethe_b}
\end{align}
where $A_i$ stands for the cardinality of set $\mathcal{A}_i$. {\color{black}As a single variable $x_i$ can be an argument of multiple factor densities $b_a(\fat{x}_a)$, its doubly counting is removed via the denominator in (\ref{Bethe_b})}. Substituting \eqref{f_fac} and (\ref{Bethe_b}) back into (\ref{F}) and exploiting the fact that $\{b_a(\fat{x}_a),b_i(x_i)\}$ are marginals of $b(\fat{x})$, {\color{black}i.e., $b_a(\fat{x}_a)= \int b(\fat{x})\mathrm{d}\fat{x}_{\backslash a}$ and $b_i(x_i)= \int b(\fat{x})\mathrm{d}\fat{x}_{\backslash i}$}, we obtain the Bethe free energy
\ifOneCol
\begin{align}
F_{\mathrm{B}}(\{b_a\},\{b_i\})=\hspace{-0cm}\sum_a\hspace{-0cm}\int\hspace{-0cm} b_a(\fat{x}_a)\ln\frac{b_a(\fat{x}_a)}{f_a(\fat{x}_a)}\mathrm{d}\fat{x}_a-\hspace{-0cm}\sum_{i}\hspace{-0cm}(A_i-1)\int b_i(x_i)\ln b_i(x_i)\mathrm{d}x_i.\label{Bethe}
\end{align}
\else
\begin{align}
F_{\mathrm{B}}(\{b_a\},\{b_i\})&=\hspace{-0cm}\sum_a\hspace{-0cm}\int\hspace{-0cm} b_a(\fat{x}_a)\ln\frac{b_a(\fat{x}_a)}{f_a(\fat{x}_a)}\mathrm{d}\fat{x}_a\notag\\
&\quad -\hspace{-0cm}\sum_{i}\hspace{-0cm}(A_i-1)\int b_i(x_i)\ln b_i(x_i)\mathrm{d}x_i.\label{Bethe}
\end{align}
\fi
The optimal solution $\hat{b}_i(x_i)$ of
\begin{align}
\min_{\{b_a\},\{b_i\}} F_{\mathrm{B}}(\{b_a\},\{b_i\})\quad \mathrm{s.t.}\, (\ref{Bethe_constr})\label{minFb1}
\end{align}
is exactly equal to $p(x_i)$ and can be found by BP~\cite{Yedidia2005}.\footnote{Since the normalization and non-negative constraints are default setting for any valid density, we will omit them for brevity.} %

If the factor graph contains cycles, we can still formulate and solve the constrained Bethe free energy minimization problem as given in (\ref{minFb1}). However, the obtained results are only approximations to $\{p(x_i)\}$. The authors of~\cite{Yedidia2005} have proven the fixed points of BP satisfy the necessary conditions for being an interior optimum (local minimum or maximum) of (\ref{minFb1}) in the discrete case. In~\cite{Heskes2003}, stable fixed points of BP were shown to be local minima. Therefore, one can regard BP as an iterative solution to the minimization problem in (\ref{minFb1}).

On top of the factor graph representation, e.g., Fig.~\ref{fig_fg}, of the target function \eqref{f_fac}, one can describe BP by specifying two rules for computing messages that are: i) from a factor node $f_a$ to a variable node $x_i$ and ii) in the reverse direction, which are respectively given as
\begin{align}
m_{a\rightarrow i}(x_i)&\propto \int f_a(\fat{x}_a)\prod_{i'\in\mathcal{I}_a\backslash i} n_{i'\rightarrow a}(x_{i'})\mathrm{d}\fat{x}_{a\backslash i} \quad \text{and}\quad 
n_{i\rightarrow a}(x_i)= \prod_{a'\in\mathcal{A}_i\backslash a} m_{a'\rightarrow i}(x_i),\label{BP_updaterule}
\end{align}
{\color{mycolor}where $\mathcal{I}_a$ stands for the index set of the entries of $\fat{x}_a$ in the complete vector $\fat{x}$. The integral in computing the 'sum-product' message $m_{a\rightarrow i}(x_i)$ represents the local marginalization associated with the factor node $f_a$ and with respect to $x_i$.} After a termination condition is satisfied, the target $f(x_i)$ is approximated by
\begin{align}
b(x_i)&\propto m_{a\rightarrow i}(x_i) n_{i\rightarrow a}(x_i)= \prod_{a'\in\mathcal{A}_i}m_{a'\rightarrow i}(x_i)\quad\quad \forall a\in\mathcal{A}_i.\label{BP_result}
\end{align}
Here we omit the specification of the normalization terms above as they can be case-dependent and often play a negligible role in computation.

Since the marginalization consistency constraint (\ref{Bethe_constr}) can often be too complicated to yield tractable messages $\{m_{a\rightarrow i}(x_i),n_{i\rightarrow a}(x_i)\}$, one natural solution is constraint relaxation, such as simplifying it to
\begin{align}
\mathrm{E}_{b_a}[\fat{t}(x_i)] = \mathrm{E}_{b_i}[\fat{t}(x_i)]\label{constrep}
\end{align}
as suggested in~\cite{Heskes2002}, where $\fat{t}(x_i)$ stands for the sufficient statistics of $x_i$ that are of concern. As shown in~\cite{Heskes2002}, solving the stationary point equations of the Bethe free energy under the constraint (\ref{constrep}) with $\fat{t}(x_i)$ from exponential family distributions yields the iterative algorithm EP characterized by the following message update rules
\begin{align}
m_{a\rightarrow i}(x_i)\propto\frac{\mathrm{Proj}_{\mathcal{Q}}\left(\hspace{-0cm}c\int f_a(\fat{x}_a)\hspace{-0cm}\prod_{i'\in\mathcal{I}_a}\hspace{-0cm}\! n_{i'\rightarrow a}(x_{i'})\mathrm{d}\fat{x}_{a\backslash i}\right)}{n_{i\rightarrow a}(x_i)}  \quad\! \text{and}\!\quad 
n_{i\rightarrow a}(x_i) =\!\!\! \prod_{a'\in\mathcal{A}_i\backslash a} \!\!\! m_{a'\rightarrow i}(x_i), \label{EP_marg}
\end{align}
where the parameter $c$ is chosen to make the argument of $\mathrm{Proj}_{\mathcal{Q}}(\cdot)$ a density of $x_i$. {\color{mycolor}Comparing (\ref{EP_marg}) with (\ref{BP_updaterule}), EP follows the same marginalization principle as BP and has one additional projection step. By limiting the a-posteriori belief in the nominator to an exponential family $\mathcal{Q}$, the messages of EP can then retain tractable forms over iterations, making it a pragmatic alternative to BP.} When considering the first- and second-order moments as the sufficient statistics to match, EP messages are Gaussian functions. 

\section{Bethe Approximation based Optimization Framework}\label{sec_framework}
The viewpoint of both BP and EP aiming to minimize Bethe free energy but under differently formalized constraints inspires us to develop a mathematical framework that permits to systematically derive message passing variants through constraint manipulation. In the following, we first show how to formulate the constraints such that VMP can be analytically attributed to the corresponding constrained Bethe free energy minimization. From a novel perspective of unifying BP, EP and VMP via constraint manipulation, we subsequently derive hybrid message passing variants in a structured manner. Finally, we introduce a modification onto the conventional factor graph for visualizing hybrid message passing. 

\subsection{VMP under constrained Bethe free energy minimization}
VMP is an iterative scheme to solve (\ref{min_mf}). We find that it is actually equivalent to adding the following constraint to the Bethe problem (\ref{minFb1}) targeted by BP
\begin{align}
b_a(\fat{x}_a) =\prod_{i\in\mathcal{I}_a} b_a(x_i)\quad \forall \, a,\label{VMP_constr}
\end{align}
where $b_a(x_i)$ is a marginal of $b_a(\fat{x}_a)$. Formally, the optimization problem (\ref{min_mf}) with VMP being a usable solution can be alternatively written as Bethe free energy minimization under both the marginalization and factorization constraint
\begin{align}
\min_{\{b_a\},\{b_i\}} F_{\mathrm{B}}(\{b_a\},\{b_i\})\quad \mathrm{s.t.}\, (\ref{Bethe_constr}),\,\, (\ref{VMP_constr}).\label{prob2}
\end{align}
The additional factorization constraint in (\ref{VMP_constr}) trivializes the marginalization consistency constraints of BP, i.e., (\ref{Bethe_constr}), and reduces the Bethe free energy to the objective function in (\ref{min_mf}). Furthermore, such factorization removes the local dependence captured by $b_a(\fat{x}_a)$, thereby turning BP into VMP.

\subsection{Hybrid VMP-BP}\label{sec:HVB}
Instead of fully factorizing $b_a(\fat{x}_a)$, we can selectively ignoring the correlation of subsets through partial factorization 
\begin{align}
b_a(\fat{x}_a) = \prod_{v} b_{a,v}(\fat{x}_{a,v}),\label{VMP_constr1}
\end{align}
where $\fat{x}_{a,v}$ is a subvector of $\fat{x}_a$, and its entries in the complete vector $\fat{x}$ are recorded by the index set $\mathcal{I}_{a,v}\subseteq\mathcal{I}_a$ that is mutually disjoint w.r.t. index sets of other subvectors of $\fat{x}_a$, i.e., $\mathcal{I}_{a,v}\cap\mathcal{I}_{a,v'\neq v}=\emptyset$. Through such constraint manipulation, we reach to a new optimization problem\looseness=-1
\begin{align}
\min_{\{b_a\},\{b_i\}} F_{\mathrm{B}}(\{b_a\},\{b_i\})\quad \mathrm{s.t.}\, (\ref{Bethe_constr})\,\, \text{and}\,\,(\ref{VMP_constr1}).\label{Bethe2}
\end{align}
As (\ref{VMP_constr1}) retains partial local dependence among the elements of $\fat{x}_{a,v}$ and we are interested in the beliefs of each element, we expect that the iterative solution to (\ref{Bethe2}) will exploit the synergy of BP and VMP. We refer to the Appendix~\ref{app:vmp-bp} for the detailed derivation and the resulting message passing rules are summarized below
\ifOneCol
\begin{align}
 m_{a\rightarrow i}(x_i)&\propto \int \mathrm{d}\fat{x}_{a,v\backslash i}{\textstyle\prod_{i'\in\mathcal{I}_{a,v\backslash i}}}n_{i'\rightarrow a}(x_{i'}) \cdot\left[ e^{\int \prod_{v'\not= v}b_{a,v'}(\fat{x}_{a,v'})\ln f_a(\fat{x}_a)\mathrm{d}\fat{x}_{a,\backslash v}}\right]\label{message1}\\
  n_{i\rightarrow a}(x_i) &= {\textstyle\prod_{a'\in\mathcal{A}_i\backslash a}} m_{a'\rightarrow i}(x_i)\label{message2}.
\end{align}
\else
\begin{align}
 m_{a\rightarrow i}(x_i)&\propto \int \mathrm{d}\fat{x}_{a,v\backslash i}\prod_{i'\in\mathcal{I}_{a,v\backslash i}}n_{i'\rightarrow a}(x_{i'}) \notag\\
&\quad\cdot\left[ e^{\int \prod_{v'\not= v}b_{a,v'}(\fat{x}_{a,v'})\ln f_a(\fat{x}_a)\mathrm{d}\fat{x}_{a,\backslash v}}\right]\label{message1}\\
  n_{i\rightarrow a}(x_i) &= \prod_{a'\in\mathcal{A}_i\backslash a} m_{a'\rightarrow i}(x_i)\label{message2}.
\end{align}
\fi
The (\ref{message1}) and (\ref{message2}) have identical forms to the message update rules of BP given in (\ref{BP_updaterule}) if we treat
\begin{align}
e^{\int \prod_{v'\not= v}b_{a,v'}(\fat{x}_{a,v'})\ln f_a(\fat{x}_a)\mathrm{d}\fat{x}_{a,\backslash v}}\label{vmp1}
\end{align}
as the factor function $f_a(\fat{x}_{a,v})$ of $\fat{x}_{a,v}$. Interestingly, the computation of the term in (\ref{vmp1}) essentially follows the rule of VMP given in (\ref{vmp_rule}). So, VMP aims at disentangling $\{\fat{x}_{a,v}\}$ and BP is responsible for computing the marginal beliefs with respect to each element in $\{\fat{x}_{a,v}\}$. This observation suggests the derived iterative solution being a hybrid VMP-BP.

Taking the factor node $f_1$ in Fig.~\ref{fig_fg} as an example, we factorize its associated belief $b_1(x_1,x_2,x_3)$ as $b_1(x_1)b_1(x_2,x_3)$. Following (\ref{message1}) and (\ref{message2}), the inner integral disentangles $x_1$ and $(x_2,x_3)$ and then $m_{a=1\rightarrow i=1}(x_1)$ boils down to VMP (the outer integral becomes trivial), whereas $x_2$ and $x_3$ are updated through BP (the outer integral removes the influence from one to the other through marginalization). From this simple example, we can see the hybrid VMP-BP permits the messages departing from the same factor node to follow different message update rules. This is a key difference to~\cite{Riegler2013} that attempts to combine VMP and BP through factor graph partitioning. From their problem formulation and also the application, the messages from the same factor node obey the same message updating rule.  For factor nodes belonging to the BP class $\mathcal{A}_{\mathrm{BP}}$, the marginalization constraints are applied to variable nodes connected to them, i.e., \cite[Equation (19)]{Riegler2013}. While for factor nodes of the mean field class $\mathcal{A}_{\mathrm{MF}}$, their beliefs are fully factorized to that of their variable nodes, making the marginalization consistency constraints trivial to fulfill. One outcome of such formulation is all messages departing from the factor nodes in $\mathcal{A}_{\mathrm{BP}}$($\mathcal{A}_{\mathrm{MF}}$) must follow BP(VMP) rule.

\subsection{Hybrid VMP-BP-EP}\label{sec:HVBE}

Very often when the variables are continuous, the marginalization consistency constraints become intractable and this issue cannot be addressed by adding partial factorization constraints. The pragmatic idea behind EP is to relax marginalization consistency into weaker moment matching constraints. In this part, we therefore apply this constraint relaxation idea onto the problem (\ref{Bethe2}) for exploiting the synergy of VMP, BP and EP. In other words, we resort to constraint manipulation, i.e., partial factorization and constraint relaxation, for easing the classic constrained Bethe free energy minimization problem.

Specifically, we divide the variable index set $\mathcal{I}$ into two disjoint index subsets, i.e., $\mathcal{I}^{[\mathrm{E}]}\cap \mathcal{I}^{[\mathrm{B}]}=\emptyset$ and $\mathcal{I}^{[\mathrm{E}]}\cup \mathcal{I}^{[\mathrm{B}]}=\mathcal{I}$. The variables under one subset $\mathcal{I}^{[\mathrm{B}]}$ are under the marginalization consistency constraints
\begin{align}\hspace{-0.4cm}
b_{a,v(i)}(x_i) = b_i(x_i)\quad \forall i\in\mathcal{I}^{[\mathrm{B}]} \,\,\forall a\in\mathcal{A}_i\label{ConstrE}.
\end{align} 
While, for others belong to $\mathcal{I}^{[\mathrm{E}]}$, their marginalization consistency constraints are relaxed into
\begin{align}\hspace{-0.2cm}
	\mathrm{E}_{b_{a,v(i)}}[\fat{t}_{a,i}(x_i)] = \mathrm{E}_{b_i}[\fat{t}_{a,i}(x_i)]\quad \forall i\in\mathcal{I}^{[\mathrm{E}]}\,\,\forall a \in\mathcal{A}_i,\label{ConstrB}
\end{align}
where $v(i)$ indexes the factor of $b_{a,v}(\fat{x}_a)$ in (\ref{VMP_constr1}) that contains $x_i$. It is noted that the sufficient statistics $\fat{t}_{a,i}$ can vary over the factor-variable node pair, i.e., being edge-dependent~\cite{discussion_Fleury}. This allows different EP variants propagating along different edges. In next Sec.~\ref{sec_sbl}, we will show even under the same sufficient statistics it would be interesting to exploit different formulations for an improved message passing performance, i.e., first- and second-order moment matching vs. mean-variance matching.

Now, our target problem becomes
\begin{align}\hspace{-0.4cm}
\min_{\{b_{a,v}\},\{b_i\}} F_{\mathrm{B}}(\{b_{a,v}\},\{b_i\})\quad \mathrm{s.t.}\, (\ref{ConstrB})\,\,\text{and}\,\, (\ref{ConstrE})\label{Bprob2}
\end{align}
and see Appendix~\ref{app:vmp-bp-ep} for deriving a hybrid VMP-BP-EP to solve it. The obtained message update rules are expressed as 
\ifOneCol
\begin{align}\hspace{-0.4cm}
m_{a\rightarrow i}(x_i) &\propto \int \mathrm{d}\fat{x}_{a,v\backslash i} \prod_{i'\in\mathcal{I}_{a,v}\backslash i}n_{i'\rightarrow a}(x_{i'}) \cdot e^{\int \prod_{v'\not= v}b_{a,v'}(\fat{x}_{a,v'})\ln f_a(\fat{x}_a)\mathrm{d}\fat{x}_{a,\backslash  v}}\quad\quad  i\in\mathcal{I}^{[\mathrm{B}]}\label{message3_0}\\
 m_{a\rightarrow i}(x_i) & \propto \frac{1}{n_{i\rightarrow a}(x_i)}\mathrm{Proj}_{\mathcal{Q}_{a,i}}\bigg[c\!\int\! \mathrm{d}\fat{x}_{a,v\backslash i}\hspace{-0.1cm}\!\prod_{i'\in\mathcal{I}_{a,v}}\!\! n_{i'\rightarrow a}(x_{i'})\cdot e^{\int \prod_{v'\not= v}b_{a,v'}(\fat{x}_{a,v'})\ln f_a(\fat{x}_a)\mathrm{d}\fat{x}_{a,\backslash  v}}\bigg]\,\, i\in\mathcal{I}^{[\mathrm{E}]}\label{message3}\\
n_{i\rightarrow a}(x_i) & = \prod_{a'\rightarrow \mathcal{A}_i\backslash a} m_{a'\rightarrow i}(x_i)\label{message3_1}
\end{align}
\else
\begin{align}
\hspace{-0.3cm} & m_{a\rightarrow i}(x_i)\notag\\
\hspace{-0.3cm} & \hspace{0.3cm} \propto \int \mathrm{d}\fat{x}_{a,v\backslash i} \prod_{i'\in\mathcal{I}_{a,v}\backslash i}n_{i'\rightarrow a}(x_{i'})\notag\\
\hspace{-0.3cm} & \hspace{0.3cm}\quad  \cdot e^{\int \prod_{v'\not= v}b_{a,v'}(\fat{x}_{a,v'})\ln f_a(\fat{x}_a)\mathrm{d}\fat{x}_{a,\backslash  v}}\quad i\in\mathcal{I}^{[\mathrm{B}]}\label{message3_0}\\
\hspace{-0.3cm} &  m_{a\rightarrow i}(x_i)\notag\\
\hspace{-0.3cm} &  \hspace{0.3cm} \propto \frac{1}{n_{i\rightarrow a}(x_i)}\mathrm{Proj}_{\mathcal{Q}_{a,i}}\bigg[c\int \mathrm{d}\fat{x}_{a,v\backslash i}\prod_{i'\in\mathcal{I}_{a,v}}n_{i'\rightarrow a}(x_{i'}).\notag\\
\hspace{-0.3cm} & \hspace{0.3cm} \quad  \cdot e^{\int \prod_{v'\not= v}b_{a,v'}(\fat{x}_{a,v'})\ln f_a(\fat{x}_a)\mathrm{d}\fat{x}_{a,\backslash  v}}\bigg]\quad i\in\mathcal{I}^{[\mathrm{E}]}\label{message3}\\
\hspace{-0.3cm} & n_{i\rightarrow a}(x_i) = \prod_{a'\rightarrow \mathcal{A}_i\backslash a} m_{a'\rightarrow i}(x_i)\label{message3_1}
\end{align}
\fi
with $c$ being a normalization constant. Comparing with (\ref{message1}) and (\ref{message2}) (i.e., hybrid VMP-BP), we only have one additional case for the message $m_{a\rightarrow i}(x_i)$ if $i\in\mathcal{I}^{[\mathrm{E}]}$, i.e., (\ref{message3}). This is mainly because for some variables the marginalization consistency constraints are relaxed to sufficient statistics matching, following the message update rule of EP. 

\ifOneCol
\begin{figure}[t!]
	\centering
	\includegraphics[scale=0.65]{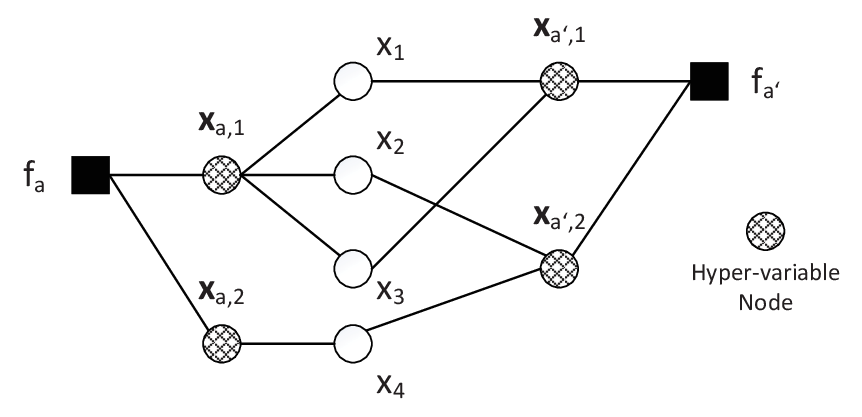}\vspace{-0.4cm}
	\caption{A modified factor graph to illustrate hybrid message passing with the factor function $f(\fat{x})=f_{a}(\fat{x})f_{a'}(\fat{x})$, the beliefs $b_a(\fat{x})=b_{a,1}(x_1,x_2,x_3)b_{a,2}(x_4)$ and $b_{a'}(\fat{x})=b_{a',1}(x_1,x_3)b_{a',2}(x_2,x_4)$.}\vspace{-0.6cm}\label{fig_vmpbp}
\end{figure}
\else
\begin{figure}[t!]
	\includegraphics[scale=0.5]{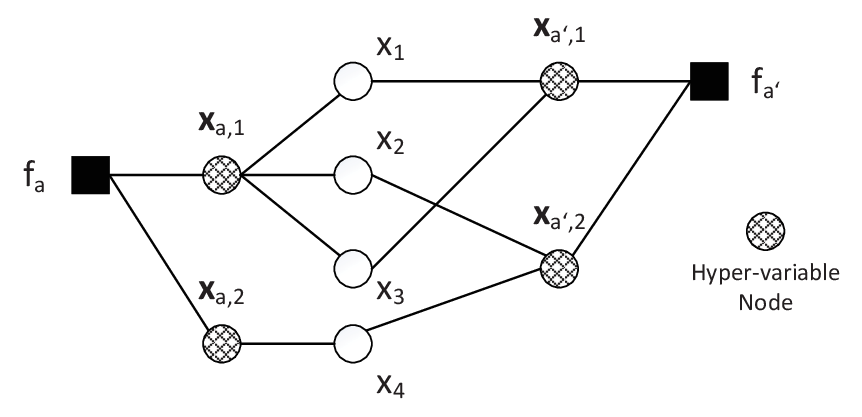}\vspace{-0cm}
	\caption{A modified factor graph to illustrate hybrid message passing with the factor function $f(\fat{x})=f_{a}(\fat{x})f_{a'}(\fat{x})$, the beliefs $b_a(\fat{x})=b_{a,1}(x_1,x_2,x_3)b_{a,2}(x_4)$ and $b_{a'}(\fat{x})=b_{a',1}(x_1,x_3)b_{a',2}(x_2,x_4)$.}\vspace{-0.2cm}\label{fig_vmpbp}
\end{figure}
\fi

\subsection{Visualization of hybrid message passing}\label{sec:VHMP}
To ease the understanding and implementation of the above-derived hybrid message passing, we propose a modification onto the factor graph, see an illustration in Fig.~\ref{fig_vmpbp}. Its key difference to the conventional factor graph, e.g., Fig.~\ref{fig_fg}, is a new type of node termed hyper-variable node. They are associated to the subvectors $\{\fat{x}_{a,v}\}$ that appear in the factorization constraints for $\{b_a(\fat{x}_a)\}$, e.g., Fig.~\ref{fig_vmpbp}. Each of them is connected to one and only one factor node. 

The message update rule for outgoing from a factor node to a hyper-variable node follows the rule of VMP, e.g.,
\begin{align}
m'_{a\rightarrow (a,v)}(\fat{x}_{a,v}) \propto e^{\int \ln f_a(\fat{x}_a) \prod_{v'\not=v} q_{a,v'}(\fat{x}_{a,v'})\mathrm{d}\fat{x}_{a,\backslash v}}\label{hybrid1}
\end{align}
with $\{q_{a,v'}\}$ given as
\begin{align}
q_{a,v'}(\fat{x}_{a,v'})= m'_{a\rightarrow (a,v')}(\fat{x}_{a,v'}) \prod_{i\in\mathcal{I}_{a,v'}} n'_{i\rightarrow (a,v')}(x_i).\label{hybrid2}
\end{align}
We can interpret $q_{a,v}(\fat{x}_{a,v})$ as the belief associated to the hyper-variable node $\fat{x}_{a,v}$. It combines the inputs from all neighboring nodes. In the language of VMP, it is also the message going to the factor node from the hyper-variable node. 

From a hyper-variable node to a normal variable node and its reverse direction, there are two cases of message updating. If the destination variable node is under the marginalization consistency constraint (i.e., $x_i$ with $i\in\mathcal{I}^{[\mathrm{B}]}$,), we take the message update rule of BP by treating the hyper-variable node as a factor node and taking the message $m'_{a\rightarrow (a,v)}(\fat{x}_{a,v})$ from the uniquely connected factor node as the factor function
\ifOneCol
\begin{align}
m'_{(a,v)\rightarrow i }(x_i)&\propto \int m'_{a\rightarrow (a,v)}(\fat{x}_{a,v}) \cdot \prod_{i'\in\mathcal{I}_{a,v} \backslash i} n'_{i'\rightarrow (a,v)}(x_{i'})\mathrm{d} \fat{x}_{a,v\backslash i},\label{hybrid3}
\end{align}
\else
\begin{align}
m'_{(a,v)\rightarrow i }(x_i)&\propto \int m'_{a\rightarrow (a,v)}(\fat{x}_{a,v})\notag\\
&\quad \cdot \prod_{i'\in\mathcal{I}_{a,v} \backslash i} n'_{i'\rightarrow (a,v)}(x_{i'})\mathrm{d} \fat{x}_{a,v\backslash i},\label{hybrid3}
\end{align}
\fi
where $n'_{i\rightarrow (a,v)}(x_i)$ given as
\begin{align}
n'_{i\rightarrow (a,v)}(x_i)=\prod_{a'\in\mathcal{A}_i\backslash a}m'_{(a',v(i))\rightarrow i }(x_i)\label{hybrid4}
\end{align}
corresponds to the message from the variable node to the hyper-variable node following the BP rule. As compared to the previously derived hybrid VMP-BP, substituting (\ref{hybrid1}) into (\ref{hybrid3}) let $m'_{(a,v)\rightarrow i }(x_i)$ become equivalent to $m_{a\rightarrow i}(x_i)$ given in (\ref{message1}). Additionally, we have the associations $n'_{i\rightarrow (a,v)}(x_i)\leftrightarrow n_{i\rightarrow a}(x_i)$ and $b_{a,v}(\fat{x}_{a,v})\leftrightarrow q_{a,v}(\fat{x}_{a,v})$.

In the other case where the destination variable node is under the moment matching constraint (i.e., $x_i$ with $i\in\mathcal{I}^{[\mathrm{E}]}$,), we shall switch to the EP rule, namely %
\ifOneCol
\begin{align}
m'_{(a,v)\rightarrow i }(x_i)&\propto \frac{1}{n'_{i\rightarrow (a,v)}(x_i)}\mathrm{Proj}_{\mathcal{Q}_i}\bigg[c\int m'_{a\rightarrow (a,v)}(\fat{x}_{a,v})\cdot \prod_{i'\in\mathcal{I}_{a,v}} n'_{i'\rightarrow (a,v)}(x_{i'})\mathrm{d} \fat{x}_{a,v\backslash i}\bigg],\label{hybrid3_2}
\end{align}
\else
\begin{align}
m'_{(a,v)\rightarrow i }(x_i)&\propto \frac{1}{n'_{i\rightarrow (a,v)}(x_i)}\mathrm{Proj}_{\mathcal{Q}_i}\left[c\int m'_{a\rightarrow (a,v)}(\fat{x}_{a,v})\right.\notag\\
&\quad \cdot \left.\prod_{i'\in\mathcal{I}_{a,v}} n'_{i'\rightarrow (a,v)}(x_{i'})\mathrm{d} \fat{x}_{a,v\backslash i}\right],\label{hybrid3_2}
\end{align}
\fi
where $n'_{i'\rightarrow (a,v)}(x_{i'})$ as the message in the reverse direction still follows (\ref{hybrid4}). Comparing with the former case, the additional $m$-projection step is the only difference here. This coincides with the known difference between EP and BP.

It is worth noting that, comparing \eqref{hybrid3}, \eqref{hybrid3_2}, \eqref{hybrid4} with \eqref{message3_0}, \eqref{message3}, \eqref{message3_1}, those equations have only slightly differences on notations. This is because $\{\mathcal{I}_{a,v}\}$ are mutually disjoint, and $m_{a\rightarrow i}(x_i)$, $n_{i\rightarrow a}(x_i)$ in Sec.~\ref{sec:HVB} and Sec.~\ref{sec:HVBE} are actually the messages passed by the hyper-variable nodes $\{\fat{x}_{a,v}\}$, i.e., denoting as $m'_{(a,v)\rightarrow i }(x_i)$ and $n'_{i'\rightarrow (a,v)}(x_{i'})$ in this part, respectively.

In short, our framework specifies the message update rules between different types of nodes on this modified factor graph, namely, defining the algorithmic structure. On top of this structure, scheduling remains as a design freedom. In principle, the order of message updating and propagating on the modified factor graph can be arbitrary, depending on applications, and may lead to different results and convergence behaviors.

\subsection{Summary}

In this section, based the framework of constrained Bethe free energy minimization, we unify three widely used message passing algorithms for a generic model. Under the same objective function (i.e., the Bethe free energy), BP, EP and VMP were associated to the marginalization consistency, moment matching and partial factorization constraints, respectively. Moment matching is weaker than marginalization consistency, but beneficial to limit the form of messages. Partial factorization permitted to ignore the variable dependencies to certain extent. From our empirical experiences on solving practical problems, we notice two motivations of performing partial factorization: 1) the random variables exhibit low dependence; and 2) it is computationally infeasible to consider the full correlation. In summary, we may interpret moment matching and partial factorization as constraint manipulation methods to trade inference fidelity for tractability.

Combining three types of constraints in a general manner, systematic derivations led us to hybrid VMP-BP-EP variants. We further mapped the corresponding message passing procedures onto a modified factor graph, visualizing hybrid message passing to assist practical uses.

\section{Application Example}\label{sec_sbl}
{\color{black}This section showcases the developed framework on solving a sparse signal recovery (SSR) problem. SSR is a classic problem that appears in different fields, e.g., active user data estimation in massive machine-type communications (mMTC) in the context of wireless communication.} In this paper, we investigate it under two problem specifications. First, assuming perfect knowledge of the statistical model, we systematically derive a new message passing algorithm that is an outcome of manipulating the constraints for deriving EP. Interestingly, it exhibits high similarity with (G)AMP, thereby being implementation friendly for large-scale systems. Next, we present a systematic derivation of hybrid message passing to efficiently solve the SSR problem without perfectly knowing the statistical model. At the end of this section, we briefly illustrate how to apply our framework for solving other practical communications problems such as massive MIMO detection and millimeter-waver channel estimation.

\subsection{EP and its variant for sparse signal recovery (SSR)}
The SSR problem of target follows a linear system model
\begin{align}
\fat{y}=\fat{H}\fat{x}+\fat{w},\label{linearssr}
\end{align}
where the channel matrix $\fat{H}\in\mathbb{C}^{N\times M}$ consisting of $N$ rows $\{\fat{h}_n\}$ represents a linear transform on the data vector $\fat{x}$, and the noise vector $\fat{w}$ has i.i.d. entries following the Gaussian distribution $\mathcal{CN}(w_n;0,\lambda^{-1})$ where $w_n$ denotes the $n$-th entry of $\fat{w}$. With i.i.d. Gaussian channel model, the entries of $\fat{H}$ follows distribution $\mathcal{CN}(h_{nm};0,N^{-1})$. The inference goal here is to estimate $\fat{x}\in\mathbb{C}^M$ based on the observation vector $\fat{y}\in\mathbb{C}^N$ and the knowledge of $\fat{H}$. Particularly, the unknown vector $\fat{x}$ represents a sparse signal only with a few non-zero entries. Here, we assume Gaussian signaling for those non-zero entries. The i.i.d. entries of $\fat{x}$ follow a Bernoulli-Gaussian distribution, i.e., $p(\fat{x}) =\prod_m (1-\rho)\delta(x_m)+\rho\mathcal{CN}(x_m;0,\alpha_m)$. This prior knowledge is critical to reliably estimate $\fat{x}$ when the system is large-scale and underdetermined ($N\ll M$). It is noted that though the considered system model in (\ref{linearssr}) is linear and additive Gaussian, the following derivation is straightforwardly extendable for generalized linear systems, i.e., the conditional density $p(\fat{y}\vert \fat{H},\fat{x})$ being arbitrary.

{\color{black}In the context of mMTC active user detection, the model assumption in above implies: The number of users outnumbers the access points, forming an underdetermined system. However, at each transmission slot, only a few users is active, i.e., $\fat{x}$ being sparse. The received signal power of each active user is denoted as $\alpha_m$. With an uplink transmit power control, one can assume $\forall m, \alpha_m=1$ without loss of generality. Besides, perfect synchronization is assumed for simplifying the discussion.}

\subsubsection{Formulation of Bethe free energy}
When the a-priori density $p(\fat{x})$ of $\fat{x}$ and the noise variance $\lambda^{-1}$ are known, our goal is to estimate $\{x_m\}$ by computing the marginals $\{p(x_m\vert\fat{y};\fat{H},\lambda)\}$ of $p(\fat{x}\vert\fat{y};\fat{H},\lambda)$. BP is not a good choice in this continuous case. Alternatively, we consider EP under the first- and second-order moment matching constraints, yielding the messages in the Gaussian family with the sufficient statistic $\fat{t}(x_i)=[\mathrm{Re}(x_i),\mathrm{Im}(x_i),\vert x_i\vert^2]^T$. 

If taking the straightforward factorization $p(\fat{x}\vert\fat{y};\fat{H},\lambda)\propto p(\fat{x})p(\fat{y}\vert\fat{x};\fat{H},\lambda)$, the resulting EP algorithm will require matrix inversion with complexity $O(N^2M)$, e.g., like S-AMP in~\cite{Cakmak2014}.\footnote{Since the factor graph of $p(\fat{x})p(\fat{y}\vert\fat{x};\fat{H},\lambda)$ is a tree, the corresponding EP is convergent and exactly yields the MMSE estimate of $\fat{x}$.} Considering the complexity issue of a large-scale system, we introduce an auxiliary vector $\fat{z}\in\mathbb{C}^N$ with the relation $\fat{z}=\fat{H}\fat{x}$. The target marginal $p(x_m\vert\fat{y};\fat{H},\lambda)$ is then alternatively proportional to the outcome of marginalizing
\begin{align}
f(\fat{z},\fat{x}) 
&=p(\fat{y}\vert\fat{z};\lambda) p(\fat{x})\prod\nolimits_{n=1}^{N}\delta(z_n-\fat{h}_n\fat{x})\label{f}
\end{align}
with respect to $x_m$. In the following, EP and its variant will be derived with respect to $f(\fat{z},\fat{x})$.\footnote{By taking $\fat{z}$ as one argument of $f(\cdot)$, the following constrained Bethe free energy minimizations will yield estimates of $\{z_n\}$ as well, even though they are not our primary goal.}%

To this end, we introduce $b_{\mathbf{z}}(\fat{z})$, $b_{\mathbf{x}}(\fat{x})$, $\{b_{\mathbf{x},\mathrm{z},n}(\fat{x},z_n)\}$ and $\{b_{\mathrm{x},m}(x_m),b_{\mathrm{z},n}(z_n)\}$ in accordance with the above factorization. It is noted that the factor function $\delta(z_n-\fat{h}_n\fat{x})$ associated to $b_{\mathbf{x},\mathrm{z},n}(\fat{x},z_n)$ is a Dirac delta function. Their KL divergence becomes infinity if there exists a support set of $b_{\mathbf{x},\mathrm{z},n}(\fat{x},z_n)$ is not a support set of $\delta(z_n-\fat{h}_n\fat{x})$, i.e.,
\ifOneCol
\begin{align}
&\int\int b_{\mathbf{x},\mathrm{z},n} (\fat{x},z_n)\ln\frac{b_{\mathbf{x},\mathrm{z},n} (\fat{x},z_n)}{\delta(z_n-\fat{h}_n\fat{x})}\mathrm{d}\fat{x}\mathrm{d}z_n
= \left\{\begin{array}{ll}
\infty &\text{if $b_{\mathbf{x},\mathrm{z},n} (z_n\vert\fat{x})\not= \delta(z_n-\fat{h}_n\fat{x})$}\\
-\mathrm{H}(b_{\mathbf{x},n}) &\text{else}\\
\end{array}\right.\label{eq:delta}
\end{align}
\else
\begin{align}
&\int\int b_{\mathbf{x},\mathrm{z},n} (\fat{x},z_n)\ln\frac{b_{\mathbf{x},\mathrm{z},n} (\fat{x},z_n)}{\delta(z_n-\fat{h}_n\fat{x})}\mathrm{d}\fat{x}\mathrm{d}z_n\notag\\
&\quad= \left\{\begin{array}{ll}
\infty &\text{if $b_{\mathbf{x},\mathrm{z},n} (z_n\vert\fat{x})\not= \delta(z_n-\fat{h}_n\fat{x})$}\\
-\mathrm{H}(b_{\mathbf{x},n}) &\text{else}\\
\end{array}\right.\label{eq:delta}
\end{align}
\fi
where $b_{\mathbf{x},\mathrm{z},n} (z_n\vert\fat{x})$ and $b_{\mathbf{x},n}(\fat{x})$ are the conditional and marginal density respectively derived from $b_{\mathbf{x},\mathrm{z},n} (\fat{x},z_n)$ according to the Bayes rule, and $\mathrm{H}(\cdot)$ stands for the entropy function. 
As the minimum of $F_{\mathrm{B}}(b)$ is of interest, it is evident to let $b_{\mathbf{x},\mathrm{z},n} (z_n,\fat{x})=b_{\mathbf{x},n}(\fat{x})\delta(z_n-\fat{h}_n\fat{x})$ so that the Bethe free energy $F_{\mathrm{B}}(b)$ can be expressed as
\ifOneCol
\begin{align}
\hspace{-0.2cm}F_{\mathrm{B}}(b)&\!=\!\KL{b_{\mathbf{x}}(\fat{x})}{p(\fat{x})}\!+\hspace{-0.2cm}\int\hspace{-0.1cm} b_{\mathbf{z}}(\fat{z})\ln\frac{b_{\mathbf{z}}(\fat{z})}{p(\fat{y}\vert\fat{z};\lambda)}\mathrm{d}\fat{z} +\hspace{-0.1cm}\sum_{m=1}^{M}N\mathrm{H}(\bi{x}{m})+\hspace{-0.1cm}\sum_{n=1}^{N}\left[\mathrm{H}(\bi{z}{n})\!-\!\mathrm{H}(b_{\mathbf{x},n})\right]. \label{Fbq}
\end{align}
\else
\begin{align}
\hspace{-0.3cm}F_{\mathrm{B}}(b)&=\KL{b_{\mathbf{x}}(\fat{x})}{p(\fat{x})}+\int b_{\mathbf{z}}(\fat{z})\ln\frac{b_{\mathbf{z}}(\fat{z})}{p(\fat{y}\vert\fat{z};\lambda)}\mathrm{d}\fat{z} \notag\\
&\quad+\sum_{m=1}^{M}N\mathrm{H}(\bi{x}{m})+\sum_{n=1}^{N}\left[\mathrm{H}(\bi{z}{n})-\mathrm{H}(b_{\mathbf{x},n})\right].\label{Fbq}
\end{align}
\fi
Conventionally, $F_{\mathrm{B}}(b)$ is minimized under the marginalization consistency constraints given as
\begin{align}
\forall m\forall n\quad \bi{x}{m}(x_m) &= \left\{\begin{array}{ll}
\int_{\fat{x}_{\backslash m}} b_{\mathbf{x},n}(\fat{x})\mathrm{d}\fat{x}_{\backslash m} & (\mathrm{a})\\
\int_{\fat{x}_{\backslash m}}b_{\mathbf{x}}(\fat{x})\mathrm{d} \fat{x}_{\backslash m} & (\mathrm{b})\\
\end{array}
\right. ;\label{consis_constx1}\\
\forall n\quad \bi{z}{n}(z_n) &= \left\{\begin{array}{ll}
\int_{\fat{x}}b_{\mathbf{x},n}(\fat{x})\delta(z_n-\fat{h}_n\fat{x})\mathrm{d} \fat{x} \\
\int_{\fat{z}_{\backslash n}}b_{\mathbf{z}}(\fat{z})\mathrm{d} \fat{z}_{\backslash n} \\
\end{array}\right. .
\label{consis_constz1}
\end{align}
\subsubsection{First- and second-order moment matching}
With a continuous random vector $\fat{x}$ of large dimension, the above constraints are often intractable. For the sake of complexity, here we relax them into the first- and second-order moment matching ones
\begin{align}
&\hspace{-0.4cm}
\begin{array}{lll}
\forall m\forall n \hspace{-0.2cm} & \mathrm{E}[x_m\vert b_{\mathrm{x},m}]=  \mathrm{E}[x_m\vert b_{\mathbf{x},n}]=\mathrm{E}[x_m\vert b_{\mathbf{x}}]&\hspace{-0.1cm} \\
\forall n \hspace{-0.2cm} & \mathrm{E}[z_n\vert b_{\mathrm{z},n}]= \mathrm{E}[\fat{h}_n\fat{x}\vert b_{\mathbf{x},n}]= \mathrm{E}[z_n\vert b_{\mathbf{z}}]&\hspace{-0.1cm} \\
\end{array};\hspace{-0.15cm}
\label{xmua}\\
&\hspace{-0.4cm}
\begin{array}{lll}
\forall m\forall n \hspace{-0.2cm}&\mathrm{E}[\vert x_m\vert^2\vert b_{\mathrm{x},m}]=\mathrm{E}[\vert x_m\vert^2\vert b_{\mathbf{x},n}]=\mathrm{E}[\vert x_m\vert^2\vert b_{\mathbf{x}}] &\hspace{-0.1cm}\\
\forall n\hspace{-0.2cm} & \mathrm{E}[\vert z_n\vert^2\vert b_{\mathrm{z},n}]=\mathrm{E}[\vert \fat{h}_n\fat{x}\vert^2\vert b_{\mathbf{x},n}]=\mathrm{E}[\vert z_n\vert^2\vert b_{\mathbf{z}}]&\hspace{-0.1cm}
\end{array}\hspace{-0.15cm}.
\label{xvar}
\end{align}
Following the method of Lagrange multipliers to minimize $F_{\mathrm{B}}(b)$ under the constraints (\ref{xmua}) and (\ref{xvar}), this yields {\color{black}an algorithm following the update rules of EP, i.e., Alg.~\ref{AlgEP}}. %

\ifOneCol
\begin{figure*}[t!]
\begin{minipage}{.48\textwidth}
\begin{algorithm}[H]
	\caption{{\color{black}EP: $\min F_{\mathrm{B}}(b)$ s.t. (\ref{xmua}) and (\ref{xvar})}}\label{AlgEP}
	\begin{algorithmic}[1]
				\State \begin{varwidth}[t]{\linewidth} \small Initialization: \par 
					$ b_{\mathbf{x}}(\fat{x})=p(\fat{x})$;\par 
					$\forall m\in\{1,2,\dots,M\},\,\,\alpha_{0,m}=\tau_{0,m}=0$;\par
					$\forall n\in\{0,1,\dots,N\},\ \forall m\in\{1,2,\dots,M\}$,\par
					$\tilde{\alpha}_{n,m}=\tilde{\tau}_{n,m}=0$.
				\end{varwidth} \vspace{0.2cm}
		\Repeat
		\State \small \begin{varwidth}[t]{\linewidth} $\forall m,$
			\par \hspace{0.5cm} $\tilde{\alpha}_{0,m}=\frac{\mathrm{E}\left[x_m\vert b_{\mathbf{x}}\right]}{\mathrm{Var}\left[x_m\vert b_{\mathbf{x}}\right]}-\alpha_{0,m}$
			\par \hspace{0.5cm} $\tilde{\tau}_{0,m}=\frac{1}{\mathrm{Var}\left[x_m\vert b_{\mathbf{x}}\right]}-\tau_{0,m}$
			\end{varwidth} \vspace{0.2cm}
			\State \small \begin{varwidth}[t]{\linewidth}  \text{For $n=1,2,\dots,N,\forall m$,}
		\par \hspace{0.5cm}	$\alpha_{n,m}=\sum_{n'=0,n'\not = n}^N\tilde{\alpha}_{n',m}$
		\par\hspace{0.5cm} $\tau_{n,m}=\sum_{n'=0,n'\not = n}^N\tilde{\tau}_{n',m}$
\end{varwidth}
		\State \begin{varwidth}[t]{\linewidth}  \text{For $n=1,2,\dots,N$,}
				\par \hspace{0.5cm} $\gamma_{\mathrm{z},n} = \sum_{m} \frac{\vert h_{nm}\vert^2}{\tau_{n,m}}$
				\par \hspace{0.5cm} $\small \mu_{\mathrm{z},n}=\sum_m  \frac{h_{nm}\alpha_{n,m}}{\tau_{n,m}}$
		\end{varwidth}	
		\State \small \begin{varwidth}[t]{\linewidth}  \text{For $n=1,2,\dots,N,\,\forall m,$} 
		\par 	\hspace{0.5cm}$\tilde{\alpha}_{n,m}=\frac{h_{nm}^*(y_n-\mu_{\mathrm{z},n})+\alpha_{n,m}\vert h_{nm}\vert^2/\tau_{n,m}}{\lambda^{-1}+\gamma_{\mathrm{z},n}-\vert h_{nm}\vert^2/\tau_{n,m}}$
		\par
		\hspace{0.5cm} $\tilde{\tau}_{n,m}=\frac{\vert h_{nm}\vert^2}{\lambda^{-1}+\gamma_{\mathrm{z},n}-\vert h_{nm}\vert^2/\tau_{n,m}}$
		\end{varwidth} \vspace{0.2cm}
		\State \begin{varwidth}[t]{\linewidth} $\forall m,$
				\par \hspace{0.5cm} $\alpha_{0,m}=\sum_{n'=1}^N\tilde{\alpha}_{n',m}$
				\par \hspace{0.5cm} $\tau_{0,m}=\sum_{n'=1}^N\tilde{\tau}_{n',m}$
		\end{varwidth}			\vspace{0.2cm}
		\State $b_{\mathbf{x}}(\fat{x})\propto p(\fat{x})\prod_m e^{-2\mathrm{Re}\left[\alpha_{0,m}^* x_m\right]+\tau_{0,m}\vert x_m\vert^2}$
		\Until{the termination condition is fulfilled.}
		\State $\hat{x}_m=  \mathrm{E}[x_m\vert b_{\mathrm{x},m}] $ 
	\end{algorithmic}
\end{algorithm} 
\vspace{0.1cm}
\end{minipage}%
\hfill
\begin{minipage}{.48\textwidth}
\begin{algorithm}[H]
	\caption{{\color{black}EP Variant: $\min F_{\mathrm{B}}(b)$ s.t. (\ref{xmua}) and (\ref{xvarb})}}\label{Alg1}
	\begin{algorithmic}[1]
		\State \begin{varwidth}[t]{\linewidth} \small Initialization: \par 
			$b_{\mathbf{x}}(\fat{x})=p(\fat{x});$ \par
			$\forall m\in\{1,2,\dots,M\},\,\, \tau_{0,m}=0$; \par
			$\forall n\!\in\!\{0,1,\dots,N\},\beta_n=0$;\par
			$\forall n\!\in\!\{0,1,\dots,N\},\forall m\!\in\!\{1,2,\dots,M\},\,\tilde{\tau}_{n,m}\!=\!0$.
			\end{varwidth}\vspace{0.2cm}
		\Repeat \small
		\State $\forall m,\,\,\tilde{\tau}_{0,m}=\frac{1}{\mathrm{Var}[x_m\vert b_{\mathbf{x}}]}-\tau_{0,m}$
		\State $\small \text{For $n=1,2,\dots,N$},\,\forall m,$\par 
		\hspace{0.5cm}$\small\tau_{n,m}=\sum_{n'=0,n'\not =n}^N \tilde{\tau}_{n',m}$
		\State $\small \text{For $n=1,2,\dots,N$},\,$\par 
		\hspace{0.5cm} $\gamma_{\mathrm{z},n} = \sum_{m} \frac{\vert h_{nm}\vert^2}{\tau_{n,m}}$
		\State $\small \text{For $n=1,2,\dots,N$},$\par
		\hspace{0.5cm} $\small\mu_{\mathrm{z},n}=\fat{h}_n\mathrm{E}[\fat{x}\vert b_{\mathbf{x}}]-\beta_{n}\gamma_{\mathrm{z},n}$
		\State $\small \text{For $n=1,2,\dots,N$},$\par
		\hspace{0.5cm} $\small \beta_n = \left(y_n-\mu_{\mathrm{z},n}\right)/(\lambda^{-1}+\gamma_{\mathrm{z},n})$	
		\State $\small \text{For $n=1,2,\dots,N$,}\,\forall m,\,$\par
		\hspace{0.5cm} $\small \tilde{\tau}_{n,m}= \frac{\vert h_{nm}\vert^2}{\gamma_{\mathrm{z},n}+\lambda^{-1}-\vert h_{nm}\vert^2/\tau_{n,m}}$
		\State $\forall m,\,\,$\par
		\hspace{0.5cm} $\tau_{0,m}=\sum_{n' =1}^N\tilde{\tau}_{n',m}$
		\State $\forall m,\,\,$\par
		\hspace{0.5cm} $\small \mu_{\mathrm{x},m}=\mathrm{E}[x_m\vert b_{\mathbf{x}}]+\tau_{0,m}^{-1}\sum_{n=1}^N h_{nm}^*\beta_n$
		\State $\small b_{\mathbf{x}}(\fat{x})\propto p(\fat{x})\prod_{m}\mathcal{CN}(x_m;\mu_{\mathrm{x},m},\tau^{-1}_{0,m})$
		\Until\small{ the termination condition is fulfilled.}		
		\State $\hat{x}_m\gets  \mathrm{E}[x_m\vert b_{\mathbf{x}}] $ 
	\end{algorithmic}
\end{algorithm}
\vspace{0.1cm}
\end{minipage}\vspace{-1.6cm}\end{figure*}
\else
\setlength{\textfloatsep}{4pt} 
\begin{algorithm}[H]
	\caption{{\color{black}EP: $\min F_{\mathrm{B}}(b)$ s.t. (\ref{xmua}) and (\ref{xvar})}}\label{AlgEP}
	\begin{algorithmic}[1]
				\State \begin{varwidth}[t]{\linewidth} \small Initialization: \par 
					$ b_{\mathbf{x}}(\fat{x})=p(\fat{x})$, $\forall m\in\{1,2,\dots,M\},\,\,\alpha_{0,m}=\tau_{0,m}=0$ \par
					$\forall n\in\{0,1,\dots,N\}\forall m\in\{1,2,\dots,M\},\,\tilde{\alpha}_{n,m}=\tilde{\tau}_{n,m}=0$ 
				\end{varwidth} \vspace{0.2cm}
		\Repeat
		\State \small \begin{varwidth}[t]{\linewidth} $\forall m,$
			\par \hspace{0.5cm} $\tilde{\alpha}_{0,m}=\frac{\mathrm{E}\left[x_m\vert b_{\mathbf{x}}\right]}{\mathrm{Var}\left[x_m\vert b_{\mathbf{x}}\right]}-\alpha_{0,m}$
			\par \hspace{0.5cm} $\tilde{\tau}_{0,m}=\frac{1}{\mathrm{Var}\left[x_m\vert b_{\mathbf{x}}\right]}-\tau_{0,m}$
			\end{varwidth} \vspace{0.2cm}
			\State \small \begin{varwidth}[t]{\linewidth}  \text{For $n=1,2,\dots,N,\forall m$,}
		\par \hspace{0.5cm}	$\alpha_{n,m}=\sum_{n'=0,n'\not = n}^N\tilde{\alpha}_{n',m}$
		\par\hspace{0.5cm} $\tau_{n,m}=\sum_{n'=0,n'\not = n}^N\tilde{\tau}_{n',m}$
\end{varwidth}
		\State \begin{varwidth}[t]{\linewidth}  \text{For $n=1,2,\dots,N$,}
				\par \hspace{0.5cm} $\gamma_{\mathrm{z},n} = \sum_{m} \frac{\vert h_{nm}\vert^2}{\tau_{n,m}}$
				\par \hspace{0.5cm} $\small \mu_{\mathrm{z},n}=\sum_m  \frac{h_{nm}\alpha_{n,m}}{\tau_{n,m}}$
		\end{varwidth}	
		\State \small \begin{varwidth}[t]{\linewidth}  \text{For $n=1,2,\dots,N,\,\forall m,$} 
		\par 	\hspace{0.5cm}$\tilde{\alpha}_{n,m}=\frac{h_{nm}^*(y_n-\mu_{\mathrm{z},n})+\alpha_{n,m}\vert h_{nm}\vert^2/\tau_{n,m}}{\lambda^{-1}+\gamma_{\mathrm{z},n}-\vert h_{nm}\vert^2/\tau_{n,m}}$
		\par
		\hspace{0.5cm} $\tilde{\tau}_{n,m}=\frac{\vert h_{nm}\vert^2}{\lambda^{-1}+\gamma_{\mathrm{z},n}-\vert h_{nm}\vert^2/\tau_{n,m}}$
		\end{varwidth} \vspace{0.2cm}
		\State \begin{varwidth}[t]{\linewidth} $\forall m,$
				\par \hspace{0.5cm} $\alpha_{0,m}=\sum_{n'=1}^N\tilde{\alpha}_{n',m}$
				\par \hspace{0.5cm} $\tau_{0,m}=\sum_{n'=1}^N\tilde{\tau}_{n',m}$
		\end{varwidth}			\vspace{0.2cm}
		\State $b_{\mathbf{x}}(\fat{x})\propto p(\fat{x})\prod_m e^{-2\mathrm{Re}\left[\alpha_{0,m}^* x_m\right]+\tau_{0,m}\vert x_m\vert^2}$
		\Until{the termination condition is fulfilled.}
		\State $\hat{x}_m=  \mathrm{E}[x_m\vert b_{\mathrm{x},m}] $ 
	\end{algorithmic}
\end{algorithm} 
\begin{algorithm}[H]
	\caption{{\color{black}EP: $\min F_{\mathrm{B}}(b)$ s.t. (\ref{xmua}) and (\ref{xvarb})}}\label{Alg1}
	\begin{algorithmic}[1]
		\State \begin{varwidth}[t]{\linewidth} \small Initialization: \par 
			$ b_{\mathbf{x}}(\fat{x})=p(\fat{x}),$ $\forall m\in\{1,2,\dots,M\},\,\, \tau_{0,m}=0$ \par
			$ \forall n\in\{0,1,\dots,N\},\beta_n=0$\par
			$\forall n\in\{0,1,\dots,N\},\forall m\in\{1,2,\dots,M\},\,\tilde{\tau}_{n,m}=0$
			\end{varwidth}
		\Repeat
		\State $\forall m,\,\,\tilde{\tau}_{0,m}=\frac{1}{\mathrm{Var}[x_m\vert b_{\mathbf{x}}]}-\tau_{0,m}$
		\State $\small \text{For $n=1,2,\dots,N$}, \forall m,\,\tau_{n,m}=\sum_{n'=0,n'\not =n}^N \tilde{\tau}_{n',m}$
		\State $\small \text{For $n=1,2,\dots,N$},\,\gamma_{\mathrm{z},n} = \sum_{m} \frac{\vert h_{nm}\vert^2}{\tau_{n,m}}$
		\State $\small \text{For $n=1,2,\dots,N$},\,\mu_{\mathrm{z},n}=\fat{h}_n\mathrm{E}[\fat{x}\vert b_{\mathbf{x}}]-\beta_{n}\gamma_{\mathrm{z},n}$
		\State $\small \text{For $n=1,2,\dots,N$},\,\beta_n = \left(y_n-\mu_{\mathrm{z},n}\right)/(\lambda^{-1}+\gamma_{\mathrm{z},n})$	
		\State $\small \text{For $n=1,2,\dots,N$,}\forall m,\,\tilde{\tau}_{n,m}= \frac{\vert h_{nm}\vert^2}{\gamma_{\mathrm{z},n}+\lambda^{-1}-\vert h_{nm}\vert^2/\tau_{n,m}}$
		\State $\forall m,\,\,\tau_{0,m}=\sum_{n' =1}^N\tilde{\tau}_{n',m}$
		\State $\forall m,\,\,\mu_{\mathrm{x},m}=\mathrm{E}[x_m\vert b_{\mathbf{x}}]+\tau_{0,m}^{-1}\sum_{n=1}^N h_{nm}^*\beta_n$
		\State $b_{\mathbf{x}}(\fat{x})\propto p(\fat{x})\prod_{m}\mathcal{CN}(x_m;\mu_{\mathrm{x},m},\tau^{-1}_{0,m})$
		\Until{the termination condition is fulfilled.}		
		\State $\hat{x}_m\gets  \mathrm{E}[x_m\vert b_{\mathbf{x}}] $ 
	\end{algorithmic}
\end{algorithm} 
\fi

{\color{black}Two remarks on the derived EP algorithms are made as follows: First, due to the notation simplification w.r.t. the specific system model, it appears differently in comparison with the classic EP presentation in terms of $\{m_{a\rightarrow i}(x_i),n_{i\rightarrow a}(x_i)\}$ and the auxiliary variable $\fat{z}$ is absorbed into the messages for $\fat{x}$.} Second, we note that from step $4$ to $6$, the computations for $n=1,2\dots,N$ are simultaneously executed. Such parallel scheduling is beneficial to limit the processing latency when $N$ tends to be large. In principle, other scheduling schemes are applicable as well. The developed framework permits to define the structure of the algorithm. Scheduling on top of it remains as a design freedom, which is beyond the scope of this work.

\subsubsection{Mean and variance consistency}
In the sequel, we illustrate how to derive a low complexity EP variant by simple constraint reformulation. Namely, under the first-order moment (mean) matching, we can equivalently and alternatively turn the second-order moment matching into the variance consistency constraint, i.e., replacing (\ref{xvar}) by 
\begin{align}
\hspace{-0.4cm}
\begin{array}{lll}
\forall m\forall n \hspace{-0.2cm}&\mathrm{Var}[x_m\vert b_{\mathrm{x},m}]=\mathrm{Var}[x_m\vert b_{\mathbf{x},n}]=\mathrm{Var}[x_m\vert b_{\mathbf{x}}] &\hspace{-0.1cm}\\
\forall n\hspace{-0.2cm} & \mathrm{Var}[z_n\vert b_{\mathrm{z},n}]=\mathrm{Var}[\fat{h}_n\fat{x}\vert b_{\mathbf{x},n}]=\mathrm{Var}[z_n\vert b_{\mathbf{z}}]&\hspace{-0.1cm}
\end{array}\hspace{-0.15cm}.
\label{xvarb}
\end{align}
By analogy, we apply the method of Lagrange multipliers for minimizing $F_{\mathrm{B}}(b)$ under (\ref{xmua}) and (\ref{xvarb}). This leads to {\color{black}the EP variant algorithm shown in} Alg.~\ref{Alg1}. Comparing it with {\color{black}the EP algorithm in} Alg.~\ref{AlgEP}, the updates for $\{\tilde{\alpha}_{n,m},\alpha_{n,m}\}$ are avoided, thereby requiring less computation efforts. Both algorithms aim at the same optimization problem as the constraints (\ref{xmua}) and (\ref{xvarb}) are equivalent to those in (\ref{xmua}) and (\ref{xvar}). %

\ifOneCol
\begin{figure*}[!t]
\begin{minipage}{.48\textwidth}
\setlength{\textfloatsep}{1pt} 
\begin{algorithm}[H]
	\caption{AMP for sparse signal recovery}\label{Alg}
	\begin{algorithmic}[1]
		\State $ \text{Initialization}:$ $\small b_{\mathbf{x}}(\fat{x})= p(\fat{x})$, $\small \forall n,\,\beta_n= 0$ 
		\Repeat \small
		\State  $\forall m,\,\,\tilde{\tau}_{0,m} =\frac{1}{\mathrm{Var}[x_m\vert b_{\mathbf{x}}]}$
		\State $\text{For $n=1,2,\dots,N$},\,\,$\par
		$ \gamma_{\mathrm{z},n}= \sum_{m=1}^{M}\vert h_{nm}\vert^2/\tilde{\tau}_{0,m}$  
		\State $\text{For $n=1,2,\dots,N$},\,\,$\par
		$ \mu_{\mathrm{z},n}= \fat{h}_n  \mathrm{E}[\fat{x}\vert b_{\mathbf{x}}] -\beta_n\gamma_{\mathrm{z},n} $
		\State $\text{For $n=1,2,\dots,N$},\,\,$\par
		$\beta_n = ( y_n-\mu_{\mathrm{z},n})/(\lambda^{-1}+\gamma_{\mathrm{z},n})$
		\State $\text{For $n=1,2,\dots,N$},\,\,\forall m ,\,\,$\par
		$\tilde{\tau}_{n,m}=\frac{\vert h_{nm}\vert^2}{\gamma_{\mathrm{z},n}+\lambda^{-1}}$
		\State $\forall m,\,\,\tau_{0,m}=\sum_{n'=1}^{N} \tilde{\tau}_{n',m} $
		\State $\forall m,\,\,$
		$\mu_{\mathrm{x},m}=  \mathrm{E}[x_m\vert b_{\mathbf{x}}] + \tau^{-1}_{0,m}\sum_n h_{nm}^*\beta_n$
		\State $b_{\mathbf{x}}(\fat{x})\propto p(\fat{x})\prod_{m=1}^{M}\mathcal{CN}(x_m;\mu_{\mathrm{x},m},\tau^{-1}_{0,m})$
		\Until{the termination condition is fulfilled.}
		\State $\hat{x}_m =  \mathrm{E}[x_m\vert b_{\mathbf{x}}] $ %
	\end{algorithmic}
\end{algorithm} %
\vspace{0.1cm}
\end{minipage}%
\hfill
\begin{minipage}{.48\textwidth}
	\setlength{\textfloatsep}{1pt}
\begin{algorithm}[H]
	\caption{{\color{black}Hybird EM-VMP-\textquotesingle EP Variant\textquotesingle \ for SBL}}\label{Alg2}
	\begin{algorithmic}[1]
		\State \begin{varwidth}[t]{\linewidth} \small Initialization: \par 
			$\forall m\in\!\{1,2,\dots,M\},\,\, \tau_{0,m}=0,\,\alpha_m = 1/M$; \par
			$\forall n\in\!\{0,1,\dots,N\},\,\,\beta_n=0$;\par
			$\forall n\in\!\{0,1,\dots,N\},\, \forall m\!\in\!\!\{1,2,\dots,M\},\,$$\tilde{\tau}_{n,m}\!=\!0$;\par
			$ b_{\mathbf{x}}(\fat{x})=\prod_m\mathcal{CN}(x_m;0,\alpha_m)$; \par
			$\lambda = \frac{100}{\mathrm{Var}(\fat{y})}$, $c=d=0,\,\, \epsilon=1.5,\,\,\eta = 1,\,\,e^{\mathrm{old}}=0$.\par
		\end{varwidth}
		\vspace{0.01cm}
		\Repeat \small
		\State $p(\fat{x})\gets \prod_{m=1}^{M}\mathcal{CN}(x_m;0,\hat{\alpha}_m)$
		\State step$\,3\sim 11$ of Alg.~\ref{Alg1}
		\State $e^{\mathrm{new}}\gets\frac{1}{N}\Vert \fat{y}-\fat{H}\hat{\fat{x}}\Vert^2 $ 
		\State ${\lambda}\gets \left[e^{\mathrm{new}}+\frac{1}{N}\sum_{n=1}^N \frac{\gamma_{\mathrm{z},n}}{1+\lambda\gamma_{\mathrm{z},n}}\right]^{-1}$
		\State $\hat{\alpha}_m\!\gets\!\{\epsilon\!-\!2\!+\!\sqrt{\!(\epsilon\!-\!2)^2\!+\!4\eta \mathrm{E}[\vert x_m\vert^2 \vert b_{\mathbf{x}}]}\}/(2\eta)$
		\State \parbox[t]{\dimexpr\linewidth-\algorithmicindent}{If $\vert e^{\mathrm{new}}-e^{\mathrm{old}}\vert < 10^{-6}$ and the support of $\mathrm{E}[\fat{x}\vert b_{\mathbf{x}}]$ is not reducing, $\epsilon\gets 0.95\epsilon$\strut}
		\State $e^{\mathrm{old}}\gets e^{\mathrm{new}}$
		\Until{the termination condition is fulfilled.}
	\end{algorithmic}
\end{algorithm}
\vspace{0.1cm}
\end{minipage}\vspace{-1.6cm}\end{figure*}
\else
\begin{algorithm}[t!]
	\caption{GAMP for sparse signal recovery}\label{Alg}
	\begin{algorithmic}[1]
		\State $\small \text{Initialization}:$ $b_{\mathbf{x}}(\fat{x})= p(\fat{x})$, $\forall n,\,\beta_n= 0$ 
		\Repeat
		\State $\forall m,\,\,\tilde{\tau}_{0,m} =\frac{1}{\mathrm{Var}[x_m\vert b_{\mathbf{x}}]}$
		\State $\small \text{For $n=1,2,\dots,N$},\,\, \gamma_{\mathrm{z},n}= \sum_{m=1}^{M}\vert h_{nm}\vert^2/\tilde{\tau}_{0,m}$  
		\State $\small \text{For $n=1,2,\dots,N$},\,\, \mu_{\mathrm{z},n}= \fat{h}_n  \mathrm{E}[\fat{x}\vert b_{\mathbf{x}}] -\beta_n\gamma_{\mathrm{z},n} $
		\State $\small \text{For $n=1,2,\dots,N$},\,\,\beta_n = ( y_n-\mu_{\mathrm{z},n})/(\lambda^{-1}+\gamma_{\mathrm{z},n})$
		\State $\small \text{For $n=1,2,\dots,N$},\,\,\forall m ,\,\,\tilde{\tau}_{n,m}=\frac{\vert h_{nm}\vert^2}{\gamma_{\mathrm{z},n}+\lambda^{-1}}$
		\State $\forall m,\,\,\tau_{0,m}=\sum_{n'=1}^{N} \tilde{\tau}_{n',m} $
		\State $\forall m,\,\,\mu_{\mathrm{x},m}=  \mathrm{E}[x_m\vert b_{\mathbf{x}}] + \tau^{-1}_{0,m}\sum_n h_{nm}^*\beta_n$
		\State $b_{\mathbf{x}}(\fat{x})\propto p(\fat{x})\prod_{m=1}^{M}\mathcal{CN}(x_m;\mu_{\mathrm{x},m},\tau^{-1}_{0,m})$
		\Until{the termination condition is fulfilled.}
		\State $\hat{x}_m =  \mathrm{E}[x_m\vert b_{\mathbf{x}}] $ %
	\end{algorithmic}
\end{algorithm} %
	\setlength{\textfloatsep}{6pt}
\begin{algorithm}[t!]
	\caption{Hybird EM-VMP-Alg.~\ref{Alg1} for SBL}\label{Alg2}
	\begin{algorithmic}[1]
		\State \begin{varwidth}[t]{\linewidth} \small Initialization: \par 
			$\forall m\in\{1,2,\dots,M\},\,\, \tau_{0,m}=0$ \par
			$ \forall n\in\{0,1,\dots,N\},\,\,\beta_n=0$\par
			$\forall n\in\{0,1,\dots,N\}\forall m\in\{1,2,\dots,M\},\,\tilde{\tau}_{n,m}=0$\par
			$\lambda = \frac{100}{\mathrm{Var}(\fat{y})},\,\,\forall m,\,\, \alpha_m = 1/M$\par
			$ b_{\mathbf{x}}(\fat{x})=\prod_m\mathcal{CN}(x_m;0,\alpha_m)$ \par
				$c=d=0,\,\, \epsilon=1.5,\,\,\eta = 1,\,\,e^{\mathrm{old}}=0$\par
		\end{varwidth}
		\Repeat
		\State $p(\fat{x})\gets \prod_{m=1}^{M}\mathcal{CN}(x_m;0,\hat{\alpha}_m)$
		\State step$\,3\sim 11$ of Alg.~\ref{Alg1}
		\State $e^{\mathrm{new}}\gets\frac{1}{N}\Vert \fat{y}-\fat{H}\hat{\fat{x}}\Vert^2 $ 
		\State ${\lambda}\gets \left[e^{\mathrm{new}}+\frac{1}{N}\sum_{n=1}^N \frac{\gamma_{\mathrm{z},n}}{1+\lambda\gamma_{\mathrm{z},n}}\right]^{-1}$
		\State $\hat{\alpha}_m\gets\{\epsilon-2+\sqrt{(\epsilon-2)^2+4\eta \mathrm{E}[\vert x_m\vert^2 \vert b_{\mathbf{x}}]}\}/(2\eta)$
		\State \parbox[t]{\dimexpr\linewidth-\algorithmicindent}{If $\vert e^{\mathrm{new}}-e^{\mathrm{old}}\vert < 10^{-6}$ and the support of $\mathrm{E}[\fat{x}\vert b_{\mathbf{x}}]$ is not reducing, $\epsilon\gets 0.95\epsilon$\strut}
		\State $e^{\mathrm{old}}\gets e^{\mathrm{new}}$
		\Until{the termination condition is fulfilled.}
	\end{algorithmic}
\end{algorithm}
\fi
	
\subsubsection{Performance comparison}
\begin{figure}[t!]
		\centering
            \includegraphics[width=1\textwidth]{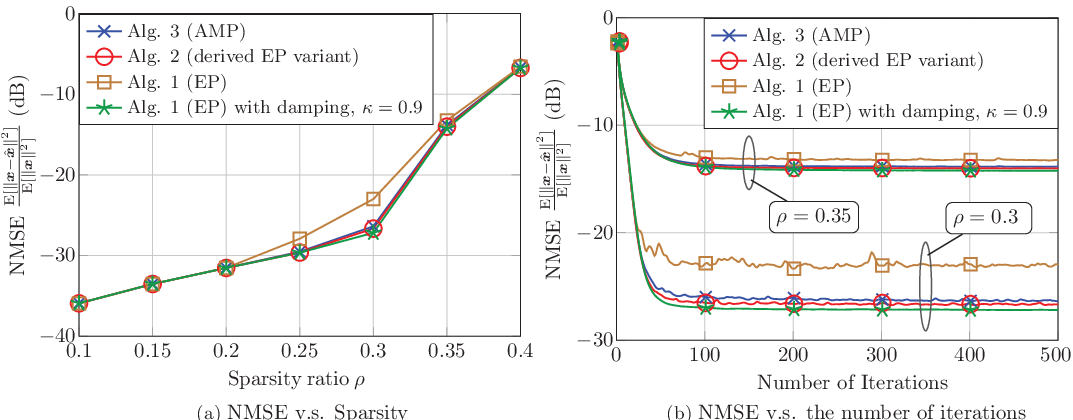}\vspace{-0.2cm}
		\caption{Normalized mean square error (NMSE) performance of message passing algorithms derived from differently constrained Bethe free energy minimization problems. In this example, the matrix $\fat{H}$ of dimension $250\times 500$ is drawn, the signal-to-noise ratio (SNR) is set to $30\,$dB and {\color{black}$\forall m, \alpha_m=1$}. 
		}\vspace{-0.6cm}\label{fig_nmse}
	\end{figure}

In this part, we compare three message passing algorithms, i.e., {\color{black}EP (Alg.~\ref{AlgEP}), EP variant (Alg.~\ref{Alg1})}, and AMP~\cite{Donoho10112009} (Alg.~\ref{Alg}) for SSR, see Fig.~\ref{fig_nmse}. 

Here, we select AMP as a benchmark algorithm due to its excellent performance attainable with low complexity in the considered system. It is shown to approximate EP in the large system limit, where the high-order terms in the message update equations are ignored~\cite{Meng2015}. It is also equivalent to S-AMP presented in~\cite{Cakmak2015} when $\fat{H}$ follows an i.i.d. zero mean Gaussian distribution. Here, we note that S-AMP in~\cite{Cakmak2015} was derived from free energy minimization under the constraint of first- and second-order moment matching {\color{black} with the auxiliary variable $\fat{z}$}. To deal with the resulting matrix inversion, it explicitly exploited the large-scale property of $\fat{H}$.

The derivations of both {\color{black}EP and EP variant algorithms} are independent of the system dimension. The key difference between {\color{black}EP variant and AMP} lies in the computation of $\{\tilde{\tau}_{n,m}\}$. Apart from that, they are nearly identical to each other, being composed of basic arithmetic operations. {\color{black} Based on this observation, we also have successfully found the constraints for deriving AMP under the framework of Bethe free energy minimization, i.e., using means and averaged variances, see Sec.~\ref{subsec:new_poing}.} GAMP~\cite{Rangan2011} generalizes AMP by adapting the step $6$ according to an arbitrarily given likelihood $p(\fat{y}\vert \fat{z})$. By analogy, we simply need to adapt the step $7$ in Alg.~\ref{Alg1} such that the resulting {\color{black}EP variant} is usable for generalized linear models as well.

Fig.~\ref{fig_nmse}~(a) shows both AMP and our proposed {\color{black}\textquotesingle EP variant\textquotesingle } achieve the best performance. On the contrary, the performance of EP (Alg.~\ref{AlgEP}) degrades severely when the sparsity ratio $\rho$ is beyond $0.25$. From our analysis, such performance degradation mainly arises from the iteration divergence under parallel message updating. This is an interesting observation, indicating that the form of constraints not only impacts the complexity and performance, but also the convergence behavior of message passing, see Fig.~\ref{fig_nmse}~(b). Therefore, it is worth to treat constraint manipulation as an important design freedom in the optimization framework when developing practical and effective message passing algorithms for various applications.

To alleviate this issue of {\color{black}EP}, we empirically introduce damping onto the step~$6$ of Alg.~\ref{AlgEP}, i.e.,
\begin{align}
\hspace{-0.5cm}
\begin{array}{l}
\tilde{\alpha}_{n,m}=(1-\kappa)\tilde{\alpha}_{n,m}+\kappa \frac{h_{nm}^*(y_n-\mu_{\mathrm{z},n})+\alpha_{n,m}\vert h_{nm}\vert^2/\tau_{n,m}}{\lambda^{-1}+\gamma_{\mathrm{z},n}-\vert h_{nm}\vert^2/\tau_{n,m}}\\
\tilde{\tau}_{n,m}=(1-\kappa)\tilde{\tau}_{n,m} + \kappa \frac{\vert h_{nm}\vert^2}{\lambda^{-1}+\gamma_{\mathrm{z},n}-\vert h_{nm}\vert^2/\tau_{n,m}}
\end{array},
\end{align}
where $\kappa\in(0,1]$ is the damping factor. In doing so the damped {\color{black}EP} is able to deliver similar performance as the others. Meanwhile, it is worth noting that the {\color{black}EP, EP variant and AMP} converge at similar speeds, and their complexities per iteration are all in the same order of $\mathcal{O}(MN)$.\footnote{For the curious readers, a more detailed analysis would show that {\color{black}EP variant and AMP} require approximately the same complexity, which is less than half of {\color{black}EP with damping}.}\looseness=-1

\subsection{Hybrid message passing for joint parameter and statistical model estimation}
For practical systems, the prior knowledge of $\lambda$ and the a-priori density $p(\fat{x})$ are in general not available, making the above-mentioned algorithms not straightforwardly applicable to estimate $\{x_m\}$. Considering the application example, the signal power of different users can be very different due to different transmit power and large-scale fading conditions, and the user activity probability $\rho$ is also in general unknown. To overcome the lack of priors, one can build models for $p(\lambda)$ and $p(\fat{x})$ such that the priors are approximately known up to certain unknown hyperparameters. By jointly estimate $\fat{x}$ and those hyperparameters, the SSR problem is extended to an empirical Bayesian estimation problem.

Before including hierarchical prior models of $\lambda$ and $\fat{x}$ into the construction of the Bethe problem, we first model the hyperprior probability density functions of signal power and the noise power as two different Gamma distributions
\begin{align}
\begin{array}{ll}
p(\alpha_m)=\mathrm{Ga}(\alpha_m\vert \epsilon,\eta); \ \ \ \ 
p(\lambda)=\mathrm{Ga}(\lambda\vert c=0,d=0),
\end{array}\notag
\end{align}
where $(\epsilon,\eta)$ and $(c,d)$ are the pre-selected shape and rate parameters of the two Gamma distributions, respectively. Next, based on the results in the work~\cite{Pedersen201594} that compares different prior models for the estimation of complex sparse signals, we choose the following parameterized Gaussian prior on $\fat{x}$
\ifOneCol
\vspace{-2mm}
\begin{align}
\begin{array}{ll}
p(\fat{x};\fGr{\hat{\alpha}})&\hspace{-0.3cm}= {\textstyle\prod_{m=1}^{M}} \mathcal{CN}(x_m;0,\hat{\alpha}_m),
\end{array}\notag
\end{align}
\else
\begin{align}
\begin{array}{ll}
p(\fat{x};\fGr{\hat{\alpha}})&\hspace{-0.3cm}=\hspace{-0.1cm}\prod\limits_{m=1}^{M}\hspace{-0.1cm}\mathcal{CN}(x_m;0,\hat{\alpha}_m),
\end{array}\notag
\end{align}
\fi
where $\fGr{\hat{\alpha}}=\{\hat{\alpha}_m\}$ denotes the estimates on hyperparameters $\{\alpha_m\}$.

From the above statistical modeling, our target function of marginalization becomes a sparse Bayesian learning (SBL) problem with parameterized Gaussian priors as
\ifOneCol
\begin{align}
 f(\fat{x},\fat{z},\lambda,\fGr{\alpha}) & = p(\fat{y}\vert\fat{z};\lambda)\cdot p(\lambda;c,d)\cdot p(\fat{x};\fGr{\alpha})\cdot {\textstyle\prod_{m=1}^{M}} p(\alpha_m;\epsilon,\eta)  {\textstyle\prod_{n=1}^{N}} \delta(z_n-\fat{h}_n\fat{x}).
\end{align}
\else
\begin{align}
 f(\fat{x},\fat{z},\lambda,\fGr{\alpha}) & = p(\fat{y}\vert\fat{z};\lambda)p(\lambda;c,d) p(\fat{x};\fGr{\alpha})\hspace{-0.1cm}\notag\\
 &\quad \cdot\prod_{m=1}^{M}\hspace{-0.1cm}p(\alpha_m;\epsilon,\eta) \hspace{-0.1cm}\prod_{n=1}^{N}\hspace{-0.1cm}\delta(z_n-\fat{h}_n\fat{x}).
\end{align}
\fi
The statistical modeling parameters $\{\lambda,\fGr{\alpha}\}$ in addition to $\{\fat{x},\fat{z}\}$ are included into the space of variational Bayesian inference. As the parameters $\{\epsilon,\eta,c,d\}$ are considered to be pre-selected, they are not taken as the arguments of $f(\cdot)$.

Following the standard way, the Bethe free energy can be written as a function of a set of densities, i.e., $b_{\mathbf{z},\lambda}(\fat{z},\lambda)$, $b_{\lambda}(\lambda)$, $b_{\mathbf{x},\fGr{\alpha}}(\fat{x},\fGr{\alpha})$, $\{b_{\alpha_m}(\alpha_m)\}$, $b_{\mathbf{x},\mathrm{z},n}(\fat{x},z_n)$ plus $\{b_{\mathrm{x},m}(x_m),b_{\mathrm{z},n}(z_n)\}$. For the sake of problem tractability, the constraints on these densities are designed to be 
\ifOneCol
\begin{align}
\begin{array}{lllllll}
b_{\mathbf{z},\lambda}(\fat{z},\lambda)&\hspace{-0.3cm}=b_{\mathbf{z}}(\fat{z})b_{\lambda}(\lambda);& \hspace{0.3cm}(a) \\
b_{\mathbf{x},\fGr{\lambda}}(\fat{x},\fGr{\alpha})&\hspace{-0.3cm}=b_{\mathbf{x}}(\fat{x})\prod_{m=1}^{M}b_{\alpha_m}(\alpha_m);&\hspace{0.3cm}(b)\\
b_{\alpha_m}(\alpha_m)&\hspace{-0.3cm}=\delta(\alpha_m-\hat{\alpha}_m)&\hspace{0.3cm}(c) &&&&
\end{array}\label{q_constr}
\end{align}
\else
\begin{align}
\begin{array}{lll}
b_{\mathbf{z},\lambda}(\fat{z},\lambda)&\hspace{-0.3cm}=b_{\mathbf{z}}(\fat{z})b_{\lambda}(\lambda);& \hspace{0.5cm}(a)\\
b_{\mathbf{x},\fGr{\lambda}}(\fat{x},\fGr{\alpha})&\hspace{-0.3cm}=b_{\mathbf{x}}(\fat{x})\prod_{m=1}^{M}b_{\alpha_m}(\alpha_m);&\hspace{0.5cm}(b)\\
b_{\alpha_m}(\alpha_m)&\hspace{-0.3cm}=\delta(\alpha_m-\hat{\alpha}_m)&\hspace{0.5cm}(c)
\end{array}\label{q_constr}
\end{align}
\fi
together with the mean and variance consistency constraints (\ref{xmua}) and (\ref{xvarb}) for the factor densities $\{b_{\mathbf{z}}(\fat{z}),b_{\mathbf{x}}(\fat{x})\}$ in relation to $\{b_{\mathrm{x},m}(x_m),b_{\mathrm{z},n}(z_n)\}$. In particular, the factorization constraints in (\ref{q_constr})-(a) and -(b) follow the idea for VMP in (\ref{VMP_constr1}). By doing so we can decouple the correlation between the model parameters $\{\fGr{\alpha},\lambda\}$ and the latent variables $\{x_m,z_n\}$. The third one, additionally letting the factor density of $\alpha_m$ be a Dirac delta-function, reduces VMP to expectation maximization (EM)~\cite{Dauwels2007}. Such a single-parameter model reduces the complexity for estimating the model parameter $\alpha_m$ at the cost of accuracy.

 	\begin{figure}[t!]
		\centering
            \includegraphics[width=\textwidth]{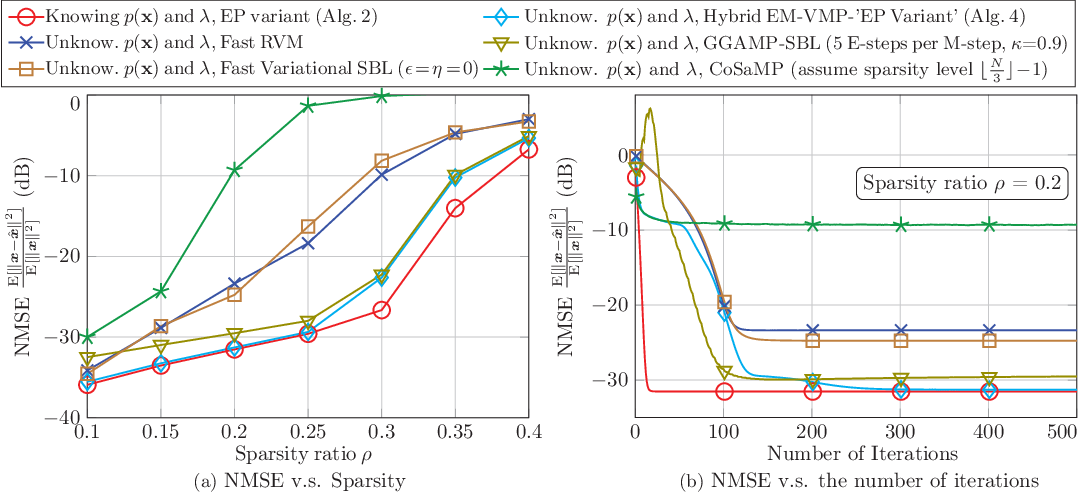}\vspace{-0.2cm}
	 		\caption{Normalized mean square error (NMSE) performance versus the sparsity ratio $\rho$ and the number of iterations. This figure is based on the same system configuration as Fig.~\ref{fig_nmse}. }\vspace{-0.6cm}\label{fig_nmse1}
	\end{figure}	
	
Applying the method of Lagrangian multipliers to minimize the Bethe free energy under the above constraints, we reach to Alg.~\ref{Alg2} that combines EM, VMP and {\color{black}EP variant}. In particular, the initialization $c=d=0$ ensures a non-informative prior for $\lambda$. The pair $(\epsilon,\eta)$ on the other hand reflects the sparsity-inducing property of the prior model on $\fat{x}$~\cite{Pedersen201594}. In particular, $\epsilon$ plays a dominant role. A smaller $\epsilon$ encourages a more sparse estimate. Without any prior knowledge of the sparsity of $\fat{x}$, we empirically propose in Alg.~\ref{Alg2} to adaptively reduce $\epsilon$ over iterations. The rationale behind the proposal follows the general idea of compressed sensing. Namely, we aim at approximating $\fat{y}$ by $\fat{H}\hat{\fat{x}}$ where the support of the estimate $\hat{\fat{x}}$ shall be as small as possible. %

Fig.~\ref{fig_nmse1}~(a) shows that the {\color{black}hybrid EM-VMP-\textquotesingle EP variant\textquotesingle }\,, i.e., Alg.~\ref{Alg2}, can approach the performance of EP variant with exact knowledge of $\lambda$ and $p(\fat{x})$ when $\fat{x}$ exhibits low and medium levels of sparsity, i.e., $\rho<0.3$. With the sparsity beyond $0.4$, the measurement ratio $\frac{N}{M}=0.5$ is too low to yield reliable estimate $\hat{\fat{x}}$. For comparison, we also include the results from the fast implementation of the relevance vector machine (RVM)~\cite{Tipping03fastmarginal}, variational SBL (i.e., VMP)~\cite{Pedersen201594}, {\color{mycolor}and the damped Gaussian GAMP-SBL~(GGAMP-SBL, i.e., a hybrid (G)AMP-EM algorithm)~\cite{GGAMP_SBL_2018}} together with a non-Bayesian approach CoSaMP~\cite{NEEDELL2009301,matlabCoSamp}. {\color{mycolor}Among these algorithms for joint parameter and statistical model estimation, the proposed message passing algorithm, i.e., Alg.~\ref{Alg2}, achieves the best NMSE for the whole range of sparsity ratio shown in the Fig.~\ref{fig_nmse1}. Further considering their convergence performance in Fig.~\ref{fig_nmse1}~(b), additional estimation of the hyper parameters cost extra iterations to converge, comparing with the case where genie knowledge on $p(\fat{x})$ and $\lambda$ is available. In order to achieve better NMSE than the others, Alg.~\ref{Alg2} takes about $300$ iterations to converge to its best NMSE. If it only needs to achieve the same NMSE as the other algorithms, Alg.~\ref{Alg2} takes similar number of iterations.} 

{\color{black}From the discussions above, we have showcased the way to apply the proposed framework under one (general) linear system model, i.e., derived several algorithms with the capability to systematically include parameter learning features. For the SSR problems under large-scale system limit, the parameter learning conventionally is applied directly through the EM algorithm \cite{Schniter2013,Rangan2014_AMP_EM}. Considering EM to be a special case of VMP, the proposed framework shows that the algorithms proposed in those works reach fixed points of differently constrained Bethe free energy minimization problems and can be derived in a systematic way. To promote this framework as a handy tool for deriving signal processing algorithms in next generation wireless communications systems, next, we will focus on its other wireless applications, e.g., data and channel estimation in spatial correlated channels, precoding, etc.  They are either already used our framework or their ad-hoc solutions fall into our framework.
}

\subsection{Application in Large-scale Wireless Communications}\label{subsec:new_poing} 

{\color{mycolor}In this part, we showcase how our framework can be applied for solving challenging receiver design problems in wireless communications. Starting from our prior work \cite{Wang2017SEU,FanHao2017SEU}, the framework was exploited for joint channel estimation and detection in massive MIMO systems. Next, we will use the framework to provide a new viewpoint for the EM-(G)AMP-Laplacian algorithm in~\cite{Yu_2019} for massive MIMO in millimeter-wave communications.}

{\color{mycolor}Our work \cite{Wang2017SEU} tackles a joint channel estimation and data detection problem for the uplink of massive MIMO systems with $N$ receive antennas and $M$ single-antenna users. As in \cite{Wang2017SEU}, we denote the vectors of the transmit signal, the received signal, the noise, and the frequency-flat channel matrix on a sub-carrier as $\fat{x}$, $\fat{y}$, $\fat{n}$, and $\fat{G}$, respectively. Under model $\fat{y=Gx+n}$ and introducing vector $\fat{z}\triangleq \fat{G}\fat{x}$, the target marginal $p(x_m|\fat{y})$ is proportional to marginalizing 
\begin{align}
p(\fat{z},\fat{y},\fat{x}, \fat{g})
&=p(\fat{y}|\fat{z}){\textstyle\prod_{n=1}^{N}} p(z_n|\fat{g}_n, \fat{x}) {\textstyle\prod_{n=1}^{N}} p(\fat{g}_n)p(\fat{x}),
\end{align}
where the $n$-th row and the $n$-th entry of $\fat{G}$ and $\fat{z}$ are denoted as $\fat{g}_n=[g_{n1},\dots,g_{nM}]$ and $z_n$, respectively. Here, the $p(z_n|\fat{g}_n, \fat{x})$ can be further written as $\delta\big(z_n-\fat{g}_n \fat{x}\big)$, and the prior distribution $p(\fat{g}_n)$ is provided by initial channel estimates. After introducing beliefs $\{b_{\mathbf{z}}(\fat{z})$, $b_{\mathbf{x,z,g},n}(\fat{x},z_n, \fat{g}_n)$, $b_{\mathbf{g},n}(\fat{g}_n)$, $ b_{\mathbf{x}}(\fat{x})\}$ and $\left\{b_{\mathrm{z},n}(z_n), b_{\mathrm{g},n,m}(g_{nm}), b_{\mathrm{x},m}(x_m)\right\}$ w.r.t. factors $\mathcal{A}=\{p(\fat{y}|\fat{z})$, $p(z_n|\fat{g}_n, \fat{x})$, $p(\fat{g}_n)$, $p(\fat{x})\}$ and variables $\mathcal{V}=\left\{z_n, g_{nm}, x_m\right\}$, respectively, the Bethe free energy is found as
\begin{align}
F_{\mathrm{B}}&(b)= \KL{b_{\mathbf{z}}(\fat{z})}{p(\fat{y}|\fat{z})}+{\textstyle\sum_n} \KL{b_{\mathbf{x,z,g},n}(\fat{x},z_n, \fat{g}_n)}{p(z_n|\fat{g}_n, \fat{x})}+{\textstyle\sum_n}\KL{b_{\mathbf{g},n}(\fat{g}_n)}{p(\fat{g}_n)}\notag\\&
+\KL{b_{\mathbf{x}}(\fat{x})}{p(\fat{x})}+{\textstyle\sum_n} \mathrm{H}[b_{\mathrm{z},n}(z_{n})]
+ {\textstyle\sum_n}{\textstyle\sum_m}\mathrm{H}[b_{\mathrm{g},n,m}(g_{nm})] +  {\textstyle\sum_m}N\mathrm{H}[b_{\mathrm{x},m}(x_m)],\label{equ:SEU_BEF}
\end{align}
where $x_m$ is the $m$-th entry of $\fat{x}$. With similar discussions to \eqref{eq:delta}, we further factorize the belief $b_{\mathbf{x,z,g},n}(\fat{x},z_n, \fat{g}_n)$ into a product of beliefs as $b_{\mathbf{x,z,g},n}(\fat{x},z_n, \fat{g}_n)= \delta(z_n-\fat{g}_n \fat{x}) q_{\mathbf{x},n}(\fat{x}) q_{\mathbf{g},n}(\fat{g}_n)$. Then, the constraints are formulated as follows: 
\begin{align}
&\begin{array}{lllllll}
\forall n \forall m \hspace{-0.2cm} & \mathrm{E}[x_m\vert b_{\mathbf{x}}]=  \mathrm{E}[x_m\vert b_{\mathrm{x},m}]= \mathrm{E}[x_m\vert q_{\mathbf{x},n}]; &\hspace{-0.1cm} \\
\forall m \  & \mathrm{Var}[x_m\vert b_{\mathrm{x}}]=\mathrm{Var}[x_m\vert b_{\mathrm{x},m}]= \frac{1}{N}\sum_n \mathrm{Var}[x_m\vert q_{\mathbf{x},n}]; &\hspace{-0.1cm} 
\end{array}\label{eq:example_AMP_x_consistant}\\
&\begin{array}{lll}
\forall n \ \hspace{0.2cm} & \mathrm{E}[z_n\vert b_{\mathbf{z}}]=\mathrm{E}[z_n\vert b_{\mathrm{z},n}]= \mathrm{E}[\fat{g}_n\fat{x}\vert q_{\mathbf{x},n}, q_{\mathbf{g},n}] ;&\hspace{-0.1cm}\\
\forall n \  \hspace{0.2cm} & \mathrm{Var}[z_n\vert b_{\mathbf{z}}]=\mathrm{Var}[z_n\vert b_{\mathrm{z},n}]= \mathrm{Var}[\fat{g}_n\fat{x}\vert q_{\mathbf{x},n}, q_{\mathbf{g},n}] ;&\hspace{-0.1cm}
\end{array} \label{eq:example_AMP_z_consistant}\\
&\begin{array}{lll}
\forall n \forall m  \hspace{-0.2cm} & \mathrm{E}[g_{mn}\vert b_{\mathbf{g},n}]=\mathrm{E}[g_{mn}\vert b_{\mathrm{g},m,n}]= \mathrm{E}[g_{mn}\vert q_{\mathbf{g},n}] ;&\hspace{-0.1cm}\\
\forall n \forall m  \hspace{-0.2cm} & \mathrm{Var}[g_{mn}\vert b_{\mathbf{g},n}]=\mathrm{Var}[g_{mn}\vert b_{\mathrm{g},m,n}]= \mathrm{Var}[g_{mn}\vert q_{\mathbf{g},n}].&\hspace{-0.1cm}
\end{array} 
\end{align}
Using fixed-point iteration to solve the stationary points of the constrained Bethe free energy, we obtain a hybrid VMP-AMP algorithm. Due to the space limit, detailed derivations, the factor graph, and the numerical evaluation over spatially correlated Rayleigh fading channels are omitted and can be found in our work \cite{Wang2017SEU}. Here, we note that the variance consistency constraints in \eqref{eq:example_AMP_x_consistant} has an averaged term over $n$, i.e., $\mathrm{Var}[x_m\vert b_{x,m}]=\frac{1}{N}\sum_{n} \mathrm{Var}[x_m\vert q_{\mathbf{x},n}]$. Namely, instead of requiring consistency for each pair $\mathrm{Var}[x_m\vert b_{\mathrm{x},m}]=\mathrm{Var}[x_m\vert q_{\mathbf{x},n}]\,\,\forall n$ as we did for deriving the EP variant Alg.~\ref{Alg1} from \eqref{xvarb}, we only ensure consistency on average in this case. This example once again shows the flexibility of constraint formulation as a design freedom under our framework.}

{\color{mycolor} The next example is not ours but from the literature, which has a heuristic combination of message passing algorithms, i.e., EM-(G)AMP-Laplacian algorithm in~\cite{Yu_2019} (Laplacian here indicates the prior distribution of angular-domain channel responses). Under our framework, we can successfully derive their algorithm as an outcome of constrained free energy minimization, therefore explain its good performance in estimating the massive MIMO millimeter-wave channel.} Using the same notations as in~\cite{Yu_2019}, we denote the vectorized angular-domain channel responses, the known coupling matrix formed by the training pilot, the received signal in the angular domain, and the noise vector as ${\fat{x}}\!\in\!\mathcal{R}^{N}$, ${\fat{A}} \!\in\! \mathcal{R}^{M\times N}$, ${\fat{y}} \!\in\! \mathcal{R}^{M}$, and ${\mathbf{w}} \!\in\! \mathcal{R}^{M}$, respectively.
Under the linear model ${\fat{y}}={\fat{A}}{\fat{x}} + {\fat{w}}$ and introducing the auxiliary vector $\fat{z}: z_m \triangleq \fat{a}_m\fat{x}$, the target marginal $p({x}_{n}|\fat{y})$ is proportional to the outcome of marginalizing
\begin{align}
f(\mathbf{z},\mathbf{x};\beta)=\prod_{m=1}^M \mathcal{N}(y_m;z_m, \sigma^2_w) \prod_{m=1}^{M} \delta\big(z_m-\fat{a}_m \fat{x}\big) \prod_{n=1}^{N} p(x_{n};\beta),\label{eq:MIMO_estimate}
\end{align}
with respect to $x_{n}$. Here, $\fat{a}_m$ is the $m$-th row vector of $\mathbf{A}$, and the $p(x_{n};\beta)=\mathrm{La}(x_{n};\beta)$ is a zero-mean Laplacian distribution with scale parameter $\beta$, i.e., $\mathrm{La}(x_{n};\beta)=e^{-|x_n|/\beta}/(2\beta)$. {\color{mycolor}After introducing beliefs $b_{\mathbf{z}}(\fat{z})$, $\{b_{\mathbf{x},\mathrm{z},m}(\fat{x}, z_m)\}$, $b_{\mathbf{x},\beta}(\fat{x},\beta)$, $b_{\beta}(\beta)$, $\{b_{x,n}(x_{n})\}$, $\{b_{\mathrm{z},m}(z_{m})\}$ and factorizing $b_{\mathbf{x},\beta}(\fat{x},\beta)$ and $b_{\mathbf{x},\mathrm{z},m}(\fat{x}, z_m)$ as $b_{\mathbf{x}}(\fat{x})b_{\beta}(\beta)$ and $q_{\mathbf{x},m}(\fat{x})\delta\big(z_m-\fat{a}_m \fat{x}\big)$, respectively, one can write the Bethe free energy as
\begin{align}
F_{\mathrm{B}}(b)= & \KL{b_{\mathbf{z}}(\fat{z})}{\prod_{m=1}^{M}\mathcal{N}(y_m;z_m,\sigma^2_w)}-\sum_{m=1}^{M}\mathrm{H}[q_{\mathbf{x},m}(\fat{x})]+\KL{b_{\mathbf{x},n}(\fat{x})b_{\beta}(\beta)}{\prod_{n=1}^{N}p(x_{n};\beta)}\notag\\&
+\sum_{n=1}^{N}M\mathrm{H}[b_{x,n}(x_{n})]
+\sum_{m=1}^{M}\mathrm{H}[b_{z,m}(z_{m})],\label{eq:BEF_Lap}
\end{align}
where the $b_{\beta}(\beta)$ is limited to its EM-estimate with $b_{\beta}(\beta)=\delta(\beta-\hat{\beta})$. The constraint set for the Bethe free energy $F_{\mathrm{B}}(b)$ in above is analogously formulated as the previous example \eqref{eq:example_AMP_x_consistant} and \eqref{eq:example_AMP_z_consistant} by flipping the notations of $m$ and $n$ and replacing $\fat{g}_n$ with perfectly known $\fat{a}_m$. Solving the problem in \eqref{eq:BEF_Lap} by the method of Lagrange multipliers, the algorithm in~\cite{Yu_2019} can be obtained.}

\section{Conclusion}\label{Sec_con}
	
This paper unified the message passing algorithms termed BP, EP and VMP under the optimization framework of constrained Bethe free energy minimization. With the same objective function, the key difference of these algorithms simply lies in the way of formulating the constraints. With this identification, it becomes natural to apply proper constraint manipulation for systematically deriving message passing variants from the same framework. In other words, one can treat constraint manipulation as a design freedom, enabling trade-offs between tractability and fidelity of approximate inference.\looseness=-1
	
In particular, we showed that a structured rather than an ad-hoc combination of BP, EP and VMP can be obtained under a set of partial factorization, marginalization consistency and moment matching constraints. Taking a classic SSR problem as an example, we subsequently derived an EP variant for SSR through constraint reformulation. It is able to outperform the standard EP algorithm with lower computational complexity. Last but not least, hybrid message passing was derived and applied for SBL when the statistical model of the system is unknown. It can deliver the performance that is close to the one attained with perfect knowledge of the statistical model. It can also outperform state-of-the-art solutions in the literature.

In short, constrained Bethe free energy minimization serves as a theoretical framework to perform joint investigation on different message passing algorithms for an improved performance. We focused on constraint manipulation in this work, illustrating its impact on the performance and complexity of message passing. For future works under this framework, it would be interesting to examine other design freedoms, such as the non-convex objective function Bethe free energy. Its construction can be influenced by, but not limited to, factorization of the target function $f(\fat{x})$, design of auxiliary variables (e.g., $\fat{z}$ in the examined SSR application), and inclusion of a temperature parameter. It is worth to note that all theses design freedoms are mutually orthogonal, implying the possibility of exploiting them in a combined manner.
	\appendices
	\numberwithin{equation}{section}
	\numberwithin{figure}{section}

\section{Derivation of hybrid VMP-BP}\label{app:vmp-bp}
Firstly, we can get rid of the constraint (\ref{VMP_constr1}) by interpreting it as a variable interchange, namely substituting $b_a(\fat{x}_a)$ with $\prod_{v} b_{a,v}(\fat{x}_{a,v})$ in the objective function $F_{\mathrm{B}}(\{b_a\},\{b_i\})$ and the other constraint (\ref{Bethe_constr}). This yields
\ifOneCol
\begin{align}
 F_{\mathrm{B}}(\{b_{a,v}\},\!\{b_i\})
&\!=\!\!\hspace{-0cm}\sum_a\hspace{-0.1cm}\!\int\!\!\hspace{-0cm} \prod_v b_{a,v}(\fat{x}_{a,v})\ln\frac{\prod_v b_{a,v}(\fat{x}_{a,v})}{f_a(\fat{x}_a)}\mathrm{d}\fat{x}_a\!-\!\hspace{-0.1cm}\sum_{i}\hspace{-0cm}\!(A_i\!-\!1)\hspace{-0.1cm} \!\int\!\! b_i(x_i)\ln b_i(x_i)\mathrm{d}x_i,\label{Bethe1}
\end{align}
\else
\begin{align}
& F_{\mathrm{B}}(\{b_{a,v}\},\{b_i\})\notag\\
&\quad=\hspace{-0cm}\sum_a\hspace{-0cm}\int\hspace{-0cm} \prod_v b_{a,v}(\fat{x}_{a,v})\ln\frac{\prod_v b_{a,v}(\fat{x}_{a,v})}{f_a(\fat{x}_a)}\mathrm{d}\fat{x}_a\notag\\
&\quad\quad -\hspace{-0cm}\sum_{i}\hspace{-0cm}(A_i-1)\int b_i(x_i)\ln b_i(x_i)\mathrm{d}x_i,\label{Bethe1}
\end{align}
\fi
while the marginalization constraint for each variable node $x_i$ in (\ref{Bethe_constr}) is simplified with the mean-field approximation $b_a(\fat{x}_a) = \prod_{v} b_{a,v}(\fat{x}_{a,v})$, i.e.,
\begin{align}
b_{a,v(i)}(x_i) = b_i(x_i)\quad \forall i \,\forall a\in\mathcal{A}_i\label{Bethe_constr1}
\end{align}
with $v(i)$ giving $i\in\mathcal{I}_{a,v(i)}$.

Secondly, we take the method of Lagrange multipliers to solve the problem
\begin{align}
\min_{\{b_{a,v}\},\{b_i\}} F_{\mathrm{B}}(\{b_{a,v}\},\{b_i\})\quad \mathrm{s.t.}\, (\ref{Bethe_constr1}).\label{Bprob1}
\end{align}
The Lagrange function is written as
\begin{align}
\hspace{-0.05cm}L_{\mathrm{B}} &= F_{\mathrm{B}}(\{b_{a,v}\},\{b_i\}) +\hspace{-0.1cm}\sum_{(a,v)}\hspace{-0.1cm}\zeta_{a,v}\left[\int b_{a,v}(\fat{x}_{a,v})\mathrm{d}\fat{x}_{a,v}-1\right] +\hspace{-0.1cm}\sum_i \hspace{-0.05cm}\zeta_i \left[\int b_{i}(x_{i})\mathrm{d}x_{i}-1\right]\notag\\
&\quad +\sum_i\hspace{-0.1cm}\sum_{a\in\mathcal{A}_i}\sum_{x_i}\lambda_{i\rightarrow a}(x_i)\left[ b_i(x_i) - b_{a,v(i)}(x_i) \right].
\end{align}
In above, we introduce the Lagrange multipliers $\{\zeta_i,\zeta_{a,v}\}$ for the implicit normalization constraints on the densities $\{b_{a,v},b_i\}$, while the additional marginalization consistency constraint (\ref{Bethe_constr1}) is associated to the Lagrange multipliers $\{\lambda_{i\rightarrow a}(x_i)\}$. Here we omit the non-negative constraints on $\{b_{a,v},b_i\}$ as later we will find that they are inherently satisfied by any interior stationary point of the Lagrange function.

Thirdly, let us now take the first-order derivatives of $L_{\mathrm{B}}$ with respect to $\{b_{a,v},b_i\}$ and $\{\lambda_{i\rightarrow a}(x_i),$ $\zeta_i,$ $\zeta_{a,v}\}$ equal to zeros. By solving the equations, we can express $\{b_{a,v},b_i\}$ as
\ifOneCol
\begin{align}
b_{a,v}(\fat{x}_{a,v})&=e^{-1-\zeta_{a,v}}\prod_{i\in\mathcal{I}_{a,v}}e^{\lambda_{i\rightarrow a}(x_i)} \cdot e^{\int \prod_{v'\not= v}b_{a,v'}(\fat{x}_{a,v'})\ln f_a(\fat{x}_a)\mathrm{d}\fat{x}_{a,\backslash  v}}\label{b_am}\\
b_i(x_i)&= e^{\frac{\zeta_i}{A_i-1}-1+\frac{1}{A_i-1}\sum_{a\in\mathcal{A}_i}\lambda_{i\rightarrow a}(x_i)},\label{b_i}
\end{align}
\else
\begin{align}
b_{a,v}(\fat{x}_{a,v})&=e^{-1-\zeta_{a,v}}\prod_{i\in\mathcal{I}_{a,v}}e^{\lambda_{i\rightarrow a}(x_i)}\notag\\
&\quad  \cdot e^{\int \prod_{v'\not= v}b_{a,v'}(\fat{x}_{a,v'})\ln f_a(\fat{x}_a)\mathrm{d}\fat{x}_{a,\backslash  v}}\label{b_am}\\
b_i(x_i)&= e^{\frac{\zeta_i}{A_i-1}-1+\frac{1}{A_i-1}\sum_{a\in\mathcal{A}_i}\lambda_{i\rightarrow a}(x_i)},\label{b_i}
\end{align}
\fi
where the Lagrange multipliers must be a solution of the following equations
\ifOneCol
\begin{align}
e^{\zeta_{a,v}+1}& = \int \mathrm{d}\fat{x}_{a,v}\prod_{i\in\mathcal{I}_{a,v}}e^{\lambda_{i\rightarrow a}(x_i)} \cdot\left[ e^{\int \prod_{v'\not=v}b_{a,v'}(\fat{x}_{a,v'})\ln f_a(\fat{x}_a)\mathrm{d}\fat{x}_{a,\backslash v}}\right],\label{lag1}\\
e^{-\frac{\zeta_i}{A_i-1}+1}&=\int  e^{\frac{1}{A_i-1}\sum_{a\in\mathcal{A}_i}\lambda_{i\rightarrow a}(x_i)}\mathrm{d}x_i,\label{lag2}\\ 
e^{\frac{1}{A_i-1}\sum_{a\in\mathcal{A}_i}\lambda_{i\rightarrow a}(x_i)}
&=e^{-\zeta_{a,v}-\frac{\zeta_i}{A_i-1}} \hspace{-0.1cm}\int\hspace{-0.1cm} \mathrm{d}\fat{x}_{a,v\backslash i} \hspace{-0.1cm} \prod_{i\in\mathcal{I}_{a,v}}\!e^{\lambda_{i\rightarrow a}(x_i)}\!\cdot\!\left[ e^{\int \prod_{v'\not=v}b_{a,v'}(\fat{x}_{a,v'})\ln f_a(\fat{x}_a)\mathrm{d}\fat{x}_{a,\backslash v}}\right].
\label{lag3}
\end{align}
\else
\begin{align}
e^{\zeta_{a,v}+1}& = \int \mathrm{d}\fat{x}_{a,v}\prod_{i\in\mathcal{I}_{a,v}}e^{\lambda_{i\rightarrow a}(x_i)} \notag\\
&\quad \cdot\left[ e^{\int \prod_{v'\not=v}b_{a,v'}(\fat{x}_{a,v'})\ln f_a(\fat{x}_a)\mathrm{d}\fat{x}_{a,\backslash v}}\right]\label{lag1}\\
e^{-\frac{\zeta_i}{A_i-1}+1}&=\int  e^{\frac{1}{A_i-1}\sum_{a\in\mathcal{A}_i}\lambda_{i\rightarrow a}(x_i)}\mathrm{d}x_i\label{lag2}\\ 
&\hspace{-1.5cm} e^{\frac{\zeta_i}{A_i-1}+\frac{1}{A_i-1}\sum_{a\in\mathcal{A}_i}\lambda_{i\rightarrow a}(x_i)}\notag\\
&\hspace{-1.5cm}\quad\quad\quad=
e^{-\zeta_{a,v}}\int \mathrm{d}\fat{x}_{a,v\backslash i}\prod_{i\in\mathcal{I}_{a,v}}e^{\lambda_{i\rightarrow a}(x_i)} \notag\\
&\hspace{-1.5cm}\quad\quad\quad\quad \cdot\left[ e^{\int \prod_{v'\not=v}b_{a,v'}(\fat{x}_{a,v'})\ln f_a(\fat{x}_a)\mathrm{d}\fat{x}_{a,\backslash v}}\right].\label{lag3}
\end{align}
\fi
As we can observe, $\{b_{a,v},b_i\}$ in the form of (\ref{b_am}) and (\ref{b_i}) are non-negative functions.

Fourthly, attempting to solve the equations of the Lagrange multipliers, the key is to determine $\{\lambda_{i\rightarrow a}(x_i)\}$ from (\ref{lag3}), while $\{\zeta_i,\zeta_{a,v}\}$ can be readily determined by them according to (\ref{lag1}) and (\ref{lag2}). For notation simplification, we introduce several auxiliary variables as
\begin{align}
\lambda_{a\rightarrow i}(x_i)&=\frac{1}{A_i-1}\sum_{a'\in\mathcal{A}_i}\lambda_{i\rightarrow a'}(x_i)-\lambda_{i\rightarrow a}(x_i) \label{var1}\\
	m_{a\rightarrow i}(x_i) &= e^{\lambda_{a\rightarrow i}(x_i)}, 
\quad\quad	 n_{i\rightarrow a}(x_i)=e^{\lambda_{i\rightarrow a}(x_i)}.\label{var3}
\end{align}
Using them, the equation (\ref{lag3}) can be alternatively written as
\ifOneCol
\begin{align}
 m_{a\rightarrow i}(x_i)&\propto \int \mathrm{d}\fat{x}_{a,v\backslash i}\prod_{i'\in\mathcal{I}_{a,v\backslash i}}n_{i'\rightarrow a}(x_{i'}) \cdot\left[ e^{\int \prod_{v'\not= v}b_{a,v'}(\fat{x}_{a,v'})\ln f_a(\fat{x}_a)\mathrm{d}\fat{x}_{a,\backslash v}}\right]\label{appmessage1}\\
  n_{i\rightarrow a}(x_i) &= \prod_{a'\in\mathcal{A}_i\backslash a} m_{a'\rightarrow i}(x_i)\label{appmessage2}.
\end{align}
\else
\begin{align}
 m_{a\rightarrow i}(x_i)&\propto \int \mathrm{d}\fat{x}_{a,v\backslash i}\prod_{i'\in\mathcal{I}_{a,v\backslash i}}n_{i'\rightarrow a}(x_{i'}) \notag\\
&\quad\cdot\left[ e^{\int \prod_{v'\not= v}b_{a,v'}(\fat{x}_{a,v'})\ln f_a(\fat{x}_a)\mathrm{d}\fat{x}_{a,\backslash v}}\right]\label{appmessage1}\\
  n_{i\rightarrow a}(x_i) &= \prod_{a'\in\mathcal{A}_i\backslash a} m_{a'\rightarrow i}(x_i)\label{appmessage2}.
\end{align}
\fi
Since $\{\mathcal{I}_{a,v}\}$ are mutually disjoint, the variable $v$ is actually $v(i)$ giving $i\in\mathcal{I}_{a,v(i)}$. The solutions to \eqref{appmessage1} and \eqref{appmessage2} yield
\ifOneCol
\begin{align}
b_{a,v}(\fat{x}_{a,v})&\propto\prod_{i\in\mathcal{I}_{a,v}} n_{i\rightarrow a}(x_i) \cdot e^{\int \prod_{v'\not= v}b_{a,v'}(\fat{x}_{a,v'})\ln f_a(\fat{x}_a)\mathrm{d}\fat{x}_{a,\backslash v}}\label{b_am1}\\
b_i(x_i)&\propto \prod_{a\in\mathcal{A}_i} m_{a\rightarrow i}(x_i),\quad \Longrightarrow\quad
b_i(x_i)\propto n_{i\rightarrow a}(x_i)m_{a\rightarrow i}(x_i)\quad \forall a \in\mathcal{A}_i.\label{b_i1}
\end{align}
\else
\begin{align}
b_{a,v}(\fat{x}_{a,v})&\propto\prod_{i\in\mathcal{I}_{a,v}} n_{i\rightarrow a}(x_i)\notag\\
&\quad  \cdot e^{\int \prod_{v'\not= v}b_{a,v'}(\fat{x}_{a,v'})\ln f_a(\fat{x}_a)\mathrm{d}\fat{x}_{a,\backslash v}}\label{b_am1}\\
b_i(x_i)&\propto \prod_{a\in\mathcal{A}_i} m_{a\rightarrow i}(x_i)\notag\\
&\propto n_{i\rightarrow a}(x_i)m_{a\rightarrow i}(x_i)\quad \forall a \in\mathcal{A}_i.\label{b_i1}
\end{align}
\fi
After normalization, they correspond to local optima of the constrained Bethe free energy. 

\section{Derivation of VMP-BP-EP}\label{app:vmp-bp-ep}
Analogously, we follow the method of Lagrange multipliers to solve the problem, starting from constructing the Lagrange function as
\ifOneCol
\begin{align}
L_{\mathrm{B}}
&= F_{\mathrm{B}}(\{b_{a,v}\},\{b_i\})\!+\!\sum_{i\in\mathcal{I}^{[\mathrm{B}]}}\sum_{a\in\mathcal{A}_i}\sum_{x_i}\lambda_{i\rightarrow a}(x_i)\left[ b_i(x_i) - b_{a,v(i)}(x_i) \right]\!+\!\hspace{-0.05cm}\sum_i \hspace{-0.05cm}\zeta_i \left[\int b_{i}(x_i)\mathrm{d}x_{i}\!-\!1\right]\notag\\
&\quad \ +\! \hspace{-0.05cm}\sum_{(a,v)}\hspace{-0.05cm}\zeta_{a,v}\left[\int b_{a,v}(\fat{x}_{a,v})\mathrm{d}\fat{x}_{a,v}\!-\!1\right] \!+\!\! \sum_{i\in\mathcal{I}^{[\mathrm{E}]}}\sum_{a\in\mathcal{A}_i} \fGr{\gamma}^T_{i\rightarrow a}\left[\mathrm{E}_{b_i}[\fat{t}_{a,i}(x_i)]\!-\!\mathrm{E}_{b_{a,v(i)}}[\fat{t}_{a,i}(x_i)] \right]  .
\end{align}
\else
\begin{align}
&\hspace{-0.2cm}L_{\mathrm{B}}\notag\\
 &= F_{\mathrm{B}}(\{b_{a,v}\},\{b_i\})+\hspace{-0.15cm}\sum_{(a,v)}\hspace{-0.05cm}\zeta_{a,v}\left[\int b_{a,v}(\fat{x}_{a,v})\mathrm{d}\fat{x}_{a,v}-1\right]   +\hspace{-0.05cm}\sum_i \hspace{-0.05cm}\zeta_i \left[\int b_{i}(x_i)\mathrm{d}x_{i}-1\right]\notag\\
&\quad + \sum_{i\in\mathcal{I}^{[\mathrm{B}]}}\sum_{a\in\mathcal{A}_i}\sum_{x_i}\lambda_{i\rightarrow a}(x_i)\left[ b_i(x_i) - b_{a,v(i)}(x_i) \right]\notag\\
&\quad + \sum_{i\in\mathcal{I}^{[\mathrm{E}]}}\sum_{a\in\mathcal{A}_i} \fGr{\gamma}^T_{i\rightarrow a}\left[\mathrm{E}_{b_i}[\fat{t}_{a,i}(x_i)]-	\mathrm{E}_{b_{a,v(i)}}[\fat{t}_{a,i}(x_i)] \right].
\end{align}
\fi
In addition to $\{\zeta_{a,v},\zeta_i,\lambda_{i\rightarrow a}(x_i)\}$, we associate the moment matching constraints with the Lagrange multipliers $\{\fGr{\gamma}_{i\rightarrow a}\}$. The dimension of each vector $\fGr{\gamma}_{i\rightarrow a}$ is identical to that of the corresponding sufficient statistic $\fat{t}_{a,i}(x_i)$.

Let us subsequently set the first-order derivatives of the Lagrange function w.r.t. the densities to zeros. In doing so, the density expressions are obtained in terms of the Lagrange multipliers, i.e.,\looseness=-1
\ifOneCol
\begin{align}
 b_{a,v}(\fat{x}_{a,v})& =e^{-1-\zeta_{a,v}}\prod_{i\in\mathcal{I}^{[\mathrm{B}]}_{a,v}}e^{\lambda_{i\rightarrow a}(x_i)}\prod_{i\in\mathcal{I}^{[\mathrm{E}]}_{a,v}}e^{\fGr{\gamma}^T_{i\rightarrow a}\fat{t}_{a,i}(x_i)} \cdot e^{\int \prod_{v'\not= v}b_{a,v'}(\fat{x}_{a,v'})\ln f_a(\fat{x}_a)\mathrm{d}\fat{x}_{a,\backslash  v}}\label{b_am1a}\\
b_i(x_i)& = e^{\frac{\zeta_i}{A_i-1}-1+\frac{1}{A_i-1}\sum_{a\in\mathcal{A}_i}\lambda_{i\rightarrow a}(x_i)},\quad \forall i\in\mathcal{I}^{[\mathrm{B}]}\label{b_i1a}\\
b_i(x_i)& = e^{\frac{\zeta_i}{A_i-1}-1+\frac{1}{A_i-1} \sum_{a\in\mathcal{A}_i}\fGr{\gamma}_{i\rightarrow a}^T\fat{t}_{a,i}(x_i)},\quad \forall i\in\mathcal{I}^{[\mathrm{E}]}\label{b_i1b},
\end{align}
\else
\begin{align}
& b_{a,v}(\fat{x}_{a,v})=e^{-1-\zeta_{a,v}}\prod_{i\in\mathcal{I}^{[\mathrm{B}]}_{a,v}}e^{\lambda_{i\rightarrow a}(x_i)}\prod_{i\in\mathcal{I}^{[\mathrm{E}]}_{a,v}}e^{\fGr{\gamma}^T_{i\rightarrow a}\fat{t}_{a,i}(x_i)}\notag\\
&\quad \quad\quad\quad\quad \cdot e^{\int \prod_{v'\not= v}b_{a,v'}(\fat{x}_{a,v'})\ln f_a(\fat{x}_a)\mathrm{d}\fat{x}_{a,\backslash  v}}\label{b_am1a}\\
& b_i(x_i)= e^{\frac{\zeta_i}{A_i-1}-1+\frac{1}{A_i-1}\sum_{a\in\mathcal{A}_i}\lambda_{i\rightarrow a}(x_i)},\quad \forall i\in\mathcal{I}^{[\mathrm{B}]}\label{b_i1a}\\
& b_i(x_i)= e^{\frac{\zeta_i}{A_i-1}-1+\frac{1}{A_i-1} \sum_{a\in\mathcal{A}_i}\fGr{\gamma}_{i\rightarrow a}^T\fat{t}_{a,i}(x_i)},\quad \forall i\in\mathcal{I}^{[\mathrm{E}]}\label{b_i1b},
\end{align}
\fi
with $\mathcal{I}^{[\mathrm{E}]}_{a,v}=\mathcal{I}_{a,v}\cap \mathcal{I}^{[\mathrm{E}]}$ and $\mathcal{I}^{[\mathrm{B}]}_{a,v}=\mathcal{I}_{a,v}\cap \mathcal{I}^{[\mathrm{B}]}$. In (\ref{b_i1b}), it is noted that $b_i(x_i)$ is a member of the exponential family $\mathcal{Q}_{a_i}$ that is characterized by the sufficient statistic $\fat{t}_{a,i}(x_i)$. 

By also letting the first-order derivatives of $L_{\mathrm{B}}$ with respect to the Lagrange multipliers be zeros, the Lagrange multipliers are constrained to ensure that: 1) the above-expressed densities are normalized to one and 2) they fulfill the constraints (\ref{ConstrB}) and (\ref{ConstrE}). In addition to the variable interchanges introduced in (\ref{var1}), here we include the following three
\begin{align}
\begin{array}{c}
\fGr{\gamma}_{a\rightarrow i} =\frac{1}{A_i-1}\sum_{a'\in\mathcal{A}_i}\fGr{\gamma}_{i\rightarrow a'}-\fGr{\gamma}_{i\rightarrow a},\\
m_{a\rightarrow i}(x_i) = \Bigg\{
\begin{array}{ll}
e^{\lambda_{a\rightarrow i}(x_i)}&\quad i\in\mathcal{I}^{[\mathrm{B}]}\\
e^{\fGr{\gamma}^T_{a\rightarrow i}\fat{t}_{a,i}(x_i)}&\quad i\in\mathcal{I}^{[\mathrm{E}]}\\
\end{array}, \quad\quad
n_{i\rightarrow a}(x_i) = \Bigg\{
\begin{array}{ll}
e^{\lambda_{i\rightarrow a}(x_i)}&\quad i\in\mathcal{I}^{[\mathrm{B}]}\\
e^{\fGr{\gamma}^T_{i\rightarrow a}\fat{t}_{a,i}(x_i)}&\quad i\in\mathcal{I}^{[\mathrm{E}]}\\
\end{array}
.
\end{array}
\end{align}
Using them, we can establish the fixed-point equations of the Lagrange multipliers as given in (\ref{message3_0}), (\ref{message3}), and (\ref{message3_1}).

	\ifCLASSOPTIONcaptionsoff
	\newpage
	\fi

	
	
	%
	\bibliographystyle{IEEEtran}
	\bibliography{IEEEabrv,DZbibfile,DZReference}

\begin{thebibliography}{10}
\providecommand{\url}[1]{#1}
\csname url@samestyle\endcsname
\providecommand{\newblock}{\relax}
\providecommand{\bibinfo}[2]{#2}
\providecommand{\BIBentrySTDinterwordspacing}{\spaceskip=0pt\relax}
\providecommand{\BIBentryALTinterwordstretchfactor}{4}
\providecommand{\BIBentryALTinterwordspacing}{\spaceskip=\fontdimen2\font plus
\BIBentryALTinterwordstretchfactor\fontdimen3\font minus
  \fontdimen4\font\relax}
\providecommand{\BIBforeignlanguage}[2]{{%
\expandafter\ifx\csname l@#1\endcsname\relax
\typeout{** WARNING: IEEEtran.bst: No hyphenation pattern has been}%
\typeout{** loaded for the language `#1'. Using the pattern for}%
\typeout{** the default language instead.}%
\else
\language=\csname l@#1\endcsname
\fi
#2}}
\providecommand{\BIBdecl}{\relax}
\BIBdecl

\bibitem{Bishop2006}
C.~M. Bishop, \emph{Pattern Recognition and Machine Learning (Information
  Science and Statistics)}.\hskip 1em plus 0.5em minus 0.4em\relax Secaucus,
  NJ, USA: Springer-Verlag New York, Inc., 2006.

\bibitem{Barber2012}
D.~Barber, \emph{Bayesian Reasoning and Machine Learning}.\hskip 1em plus 0.5em
  minus 0.4em\relax New York, NY, USA: Cambridge University Press, 2012.

\bibitem{Jordan1999}
M.~I. Jordan, Z.~Ghahramani, T.~S. Jaakkola, and L.~K. Saul, ``An introduction
  to variational methods for graphical models,'' \emph{Mach. Learn.}, vol.~37,
  no.~2, pp. 183--233, Nov. 1999.

\bibitem{Yedidia2005}
J.~Yedidia, W.~Freeman, and Y.~Weiss, ``Constructing free-energy approximations
  and generalized belief propagation algorithms,'' \emph{{IEEE Trans. Inf.
  Theory}}, vol.~51, no.~7, pp. 2282--2312, Jul. 2005.

\bibitem{Winn05}
J.~Winn, C.~Bishop, and T.~Jaakkola, ``Variational message passing,'' \emph{J.
  of Machine Learning Research}, vol.~6, pp. 661--694, Apr. 2005.

\bibitem{Dempster77}
A.~P. Dempster, N.~M. Laird, and D.~B. Rubin, ``Maximum likelihood from
  incomplete data via the {EM} algorithm,'' \emph{J. the Royal Statistical
  Society. Series B}, vol.~39, no.~1, pp. 1--38, 1977.

\bibitem{Pearl}
J.~Pearl, \emph{Probabilistic Reasoning in Intelligent Systems: Networks of
  Plausible Inference}.\hskip 1em plus 0.5em minus 0.4em\relax San Francisco,
  CA, USA: Morgan Kaufmann Publishers Inc., 1988.

\bibitem{Minka2001}
T.~P. Minka, ``Expectation propagation for approximate {Bayesian} inference,''
  in \emph{Proc. of Conf. on Uncertainty in Artificial Intelligence}, San
  Francisco, CA, USA, 2001, pp. 362--369.

\bibitem{LLiu2019}
L.~{Liu}, C.~{Yuen}, Y.~L. {Guan}, Y.~{Li}, and C.~{Huang}, ``Gaussian message
  passing for overloaded massive mimo-noma,'' \emph{{IEEE Trans. Wireless
  Commun.}}, vol.~18, no.~1, pp. 210--226, 2019.

\bibitem{LLiu2019_b}
L.~{Liu}, Y.~{Chi}, C.~{Yuen}, Y.~L. {Guan}, and Y.~{Li}, ``Capacity-achieving
  mimo-noma: Iterative lmmse detection,'' \emph{{IEEE Trans. Signal Process.}},
  vol.~67, no.~7, pp. 1758--1773, 2019.

\bibitem{Zhang2016SPL}
D.~{Zhang}, M.~{Matth\'e}, L.~L. {Mendes}, and G.~{Fettweis}, ``Message passing
  algorithms for upper and lower bounding the coded modulation capacity in a
  large-scale linear system,'' \emph{IEEE Sig. Process. Lett.}, vol.~23, no.~4,
  pp. 537--540, Apr. 2016.

\bibitem{Hansen_Fleury_2018}
T.~L. {Hansen}, P.~B. {J{\o}rgensen}, M.~{Badiu}, and B.~H. {Fleury}, ``An
  iterative receiver for ofdm with sparsity-based parametric channel
  estimation,'' \emph{{IEEE Trans. Signal Process.}}, vol.~66, no.~20, pp.
  5454--5469, Oct 2018.

\bibitem{Heath_2019}
N.~J. {Myers} and R.~W. {Heath}, ``Message passing-based joint cfo and channel
  estimation in mmwave systems with one-bit adcs,'' \emph{{IEEE Trans. Wireless
  Commun.}}, vol.~18, no.~6, pp. 3064--3077, June 2019.

\bibitem{Donoho10112009}
D.~L. Donoho, A.~Maleki, and A.~Montanari, ``Message-passing algorithms for
  compressed sensing,'' \emph{Proc. of the National Academy of Sciences}, vol.
  106, no.~45, pp. 18\,914--18\,919, Oct. 2009.

\bibitem{Rangan2011}
S.~Rangan, ``Generalized approximate message passing for estimation with random
  linear mixing,'' in \emph{Proc. IEEE Inf. Symp. Inf. Theory}, Saint
  Petersburg, Russia, Jul. 2011, pp. 2168--2172.

\bibitem{Cakmak2014}
B.~\c{C}akmak, O.~Winther, and B.~H. Fleury, ``{S-AMP}: Approximate message
  passing for general matrix ensembles,'' in \emph{Proc. IEEE Inf. Theory
  Workshop}, Hobart, Tasmania, Australia, Nov. 2014, pp. 192--196.

\bibitem{Meng2015}
X.~Meng, S.~Wu, L.~Kuang, and J.~Lu, ``An expectation propagation perspective
  on approximate message passing,'' \emph{{IEEE Signal Processing Lett.}},
  vol.~22, no.~8, pp. 1194--1197, Aug. 2015.

\bibitem{new_insight_GAMP}
L.~{Liu}, Y.~{Li}, C.~{Huang}, C.~{Yuen}, and Y.~L. {Guan}, ``A new insight
  into gamp and amp,'' \emph{{IEEE Trans. Veh. Technol.}}, vol.~68, no.~8, pp.
  8264--8269, 2019.

\bibitem{sWu2014}
S.~Wu \emph{et~al.}, ``Low-complexity iterative detection for large-scale
  multiuser {MIMO-OFDM} systems using approximate message passing,''
  \emph{{IEEE J. Sel. Areas Signal Process.}}, vol.~8, no.~5, pp. 902--915,
  Oct. 2014.

\bibitem{sWang2015a}
S.~Wang \emph{et~al.}, ``Energy-efficient and low-complexity uplink transceiver
  for massive spatial modulation {MIMO},'' \emph{{IEEE Trans. Veh. Technol.}},
  vol.~64, no.~10, pp. 4617--4632, Oct. 2015.

\bibitem{Lyu_2019}
S.~{Lyu} and C.~{Ling}, ``Hybrid vector perturbation precoding: The blessing of
  approximate message passing,'' \emph{{IEEE Trans. Signal Process.}}, vol.~67,
  no.~1, pp. 178--193, Jan 2019.

\bibitem{Jeon2015}
C.~Jeon \emph{et~al.}, ``Optimality of large {MIMO} detection via approximate
  message passing,'' in \emph{Proc. IEEE Int. Symp. on Inf. Theory}, Hong Kong,
  China, Jun. 2015, pp. 1227--1231.

\bibitem{capacity_AMP_CM}
L.~{Liu}, C.~{Liang}, J.~{Ma}, and P.~{Li}, ``Capacity optimality of amp in
  coded systems,'' \emph{submitted to IEEE Trans. Inf. Theory}, 2019, available
  online at arXiv:1901.09559.

\bibitem{VAMP_2017}
S.~{Rangan}, P.~{Schniter}, and A.~K. {Fletcher}, ``Vector approximate message
  passing,'' in \emph{Proc. IEEE Int. Symp. on Inf. Theory}, 2017, pp.
  1588--1592.

\bibitem{GVAMP_2016}
P.~{Schniter}, S.~{Rangan}, and A.~K. {Fletcher}, ``Vector approximate message
  passing for the generalized linear model,'' in \emph{Asilomar Conf. on
  Signals, Syst. and Comput.}, 2016, pp. 1525--1529.

\bibitem{OAMP_2017}
J.~{Ma} and L.~{Ping}, ``Orthogonal amp,'' \emph{IEEE Access}, vol.~5, pp.
  2020--2033, 2017.

\bibitem{GrAMP_2018}
X.~{Meng}, S.~{Wu}, and J.~{Zhu}, ``A unified bayesian inference framework for
  generalized linear models,'' \emph{{IEEE Signal Process. Lett.}}, vol.~25,
  no.~3, pp. 398--402, 2018.

\bibitem{FanHao2017SEU}
H.~{Fan}, W.~{Wang}, D.~{Zhang}, and X.~{Gao}, ``Generalized approximate
  message passing detection with row-orthogonal linear preprocessing for uplink
  massive mimo systems,'' in \emph{Proc. IEEE Veh. Technol. Conf.}, 2017, pp.
  1--6.

\bibitem{Shutin2011}
D.~Shutin, T.~Buchgraber, S.~Kulkarni, and H.~Poor, ``Fast variational sparse
  {Bayesian} learning with automatic relevance determination for superimposed
  signals,'' \emph{{IEEE Trans. Signal Process.}}, vol.~59, no.~12, pp.
  6257--6261, Dec. 2011.

\bibitem{Schniter2013}
J.~P. {Vila} and P.~{Schniter}, ``Expectation-maximization gaussian-mixture
  approximate message passing,'' \emph{{IEEE Trans. Signal Process.}}, vol.~61,
  no.~19, pp. 4658--4672, 2013.

\bibitem{Rangan2012_AMP_EM}
U.~Kamilov, S.~Rangan, M.~Unser, and A.~K. Fletcher, ``Approximate message
  passing with consistent parameter estimation and applications to sparse
  learning,'' in \emph{Advances in Neural Info. Process. Systems}, 2012, pp.
  2438--2446.

\bibitem{GGAMP_SBL_2018}
M.~{Al-Shoukairi}, P.~{Schniter}, and B.~D. {Rao}, ``A gamp-based low
  complexity sparse bayesian learning algorithm,'' \emph{{IEEE Trans. Signal
  Process.}}, vol.~66, no.~2, pp. 294--308, 2018.

\bibitem{Utkovski2017}
Z.~Utkovski \emph{et~al.}, ``Random access in {C-RAN} for user activity
  detection with limited-capacity fronthaul,'' \emph{{IEEE Signal Process.
  Lett.}}, vol.~24, no.~1, pp. 17--21, Jan. 2017.

\bibitem{mutual_broadcast}
L.~{Zhang} and D.~{Guo}, ``Wireless peer-to-peer mutual broadcast via sparse
  recovery,'' in \emph{Proc. IEEE Int. Symp. on Inf. Theory}, 2011, pp.
  1901--1905.

\bibitem{Mo2014}
J.~Mo \emph{et~al.}, ``Channel estimation in millimeter wave {MIMO} systems
  with one-bit quantization,'' in \emph{Asilomar Conf. on Signals, Syst. and
  Comput.}, Pacific Grove, CA, USA, Nov. 2014, pp. 957--961.

\bibitem{Karseras2015}
E.~Karseras \emph{et~al.}, ``Fast variational {Bayesian} learning for channel
  estimation with prior statistical information,'' in \emph{Proc. IEEE Int.
  Workshop on Signal Process. Adv. in Wireless Commun.}, Stockholm, Sweden,
  Jun. 2015, pp. 470--474.

\bibitem{Stockle2016}
C.~St\"ockle \emph{et~al.}, ``Channel estimation in massive {MIMO} systems
  using 1-bit quantization,'' in \emph{Proc. IEEE Int. Workshop on Signal
  Process. Adv. in Wireless Commun.}, Jul. 2016, pp. 1--6.

\bibitem{Wu_EM_AMP_16}
X.~{Meng}, S.~{Wu}, L.~{Kuang}, D.~{Huang}, and J.~{Lu}, ``Multi-user detection
  for spatial modulation via structured approximate message passing,''
  \emph{{IEEE Commun. Lett.}}, vol.~20, no.~8, pp. 1527--1530, 2016.

\bibitem{Joint_channel_user_activity_one}
L.~{Liu} and W.~{Yu}, ``Massive connectivity with massive mimo -{Part I}:
  Device activity detection and channel estimation,'' \emph{{IEEE Trans. Signal
  Process.}}, vol.~66, no.~11, pp. 2933--2946, 2018.

\bibitem{Joint_channel_user_activity_two}
------, ``Massive connectivity with massive mimo -{Part II}: Achievable rate
  characterization,'' \emph{{IEEE Trans. Signal Process.}}, vol.~66, no.~11,
  pp. 2947--2959, 2018.

\bibitem{Luo_UTAMP}
M.~{Luo}, Q.~{Guo}, D.~{Huang}, and J.~{Xi}, ``Sparse bayesian learning based
  on approximate message passing with unitary transformation,'' in \emph{Proc.
  IEEE VTS Asia Pacific Wireless Commun. Sym.}, 2019, pp. 1--5.

\bibitem{Senst2011a}
M.~Senst and G.~Ascheid, ``A combined belief propagation and mean field
  algorithm for soft carrier phase estimation,'' in \emph{Proc. IEEE Inter.
  Symp. on Wireless Commun. Syst.}, Aachen, Germany, Nov. 2011, pp. 512--516.

\bibitem{Badiu2012}
M.~A. Badiu \emph{et~al.}, ``Message-passing algorithms for channel estimation
  and decoding using approximate inference,'' in \emph{Proc. IEEE Inter. Symp.
  on Inf. Theory}, MA, USA, Jul. 2012, pp. 2376--2380.

\bibitem{Sun2015}
P.~Sun, C.~Zhang, Z.~Wang, C.~Manchon, and B.~Fleury, ``Iterative receiver
  design for {ISI} channels using combined belief- and
  expectation-propagation,'' \emph{{IEEE Signal Processing Lett.}}, vol.~22,
  no.~10, pp. 1733--1737, Oct. 2015.

\bibitem{wWang2016}
W.~Wang, Z.~Wang, C.~Zhang, Q.~Guo, P.~Sun, and X.~Wang, ``A {BP-MF-EP} based
  iterative receiver for joint phase noise estimation, equalization, and
  decoding,'' \emph{{IEEE Signal Processing Lett.}}, vol.~23, no.~10, pp.
  1349--1353, Oct. 2016.

\bibitem{NWU2017}
N.~Wu, W.~Yuan, Q.~Guo, and J.~Kuang, ``A hybrid {BP-EP-VMP} approach to joint
  channel estimation and decoding for {FTN} signaling over frequency selective
  fading channels,'' \emph{IEEE Access}, vol.~5, pp. 6849--6858, May 2017.

\bibitem{Riegler2013}
E.~Riegler, G.~Kirkelund, C.~Manchon, M.~Badiu, and B.~Fleury, ``Merging belief
  propagation and the mean field approximation: A free energy approach,''
  \emph{{IEEE Trans. Inf. Theory}}, vol.~59, no.~1, pp. 588--602, Jan. 2013.

\bibitem{Wang2017SEU}
S.~{Chen}, W.~{Wang}, D.~{Zhang}, and X.~{Gao}, ``Robust approximate message
  passing detection based on minimizing bethe free energy for massive mimo
  systems,'' in \emph{Proc. IEEE Veh. Technol. Conf.}, Sep. 2017, pp. 1--6.

\bibitem{Kschischang2001}
F.~R. Kschischang, B.~J. Frey, and H.~A. Loeliger, ``Factor graphs and the
  sum-product algorithm,'' \emph{{IEEE Trans. Inf. Theory}}, vol.~47, no.~2,
  pp. 498--519, Feb. 2001.

\bibitem{Heskes2003}
T.~Heskes, ``Stable fixed points of loopy belief propagation are minima of the
  {B}ethe free energy,'' in \emph{Proc. Neural Inf. Process. Syst.}, Nevada,
  USA, 2003.

\bibitem{Heskes2002}
T.~Heskes and O.~Zoeter, ``Expectation propagation for approximate inference in
  dynamic {B}ayesian networks,'' in \emph{Proc. Conf. on Uncertain. in Artif.
  Intell.}, Alberta, Canada, 2002, pp. 216--223.

\bibitem{discussion_Fleury}
The idea in generalizing the constraints to be edge-dependent comes from our
  private correspondence with Bernard H. Fleury, Aalborg University, Denmark.

\bibitem{Cakmak2015}
B.~\c{C}akmak, O.~Winther, and B.~H. Fleury, ``S-amp for non-linear observation
  models,'' in \emph{Proc. IEEE Inf. Symp. Inf. Theory}, June 2015, pp.
  2807--2811.

\bibitem{Pedersen201594}
N.~L. Pedersen, C.~N. Manch\'on, M.-A. Badiu, D.~Shutin, and B.~H. Fleury,
  ``Sparse estimation using {Bayesian} hierarchical prior modeling for real and
  complex linear models,'' \emph{Signal Processing}, vol. 115, pp. 94--109,
  Mar. 2015.

\bibitem{Dauwels2007}
J.~Dauwels, ``On variational message passing on factor graphs,'' in \emph{Proc.
  IEEE Inf. Symp. Inf. Theory}, Nice, France, Jun. 2007, pp. 2546--2550.

\bibitem{Tipping03fastmarginal}
M.~E. Tipping and A.~Faul, ``Fast marginal likelihood maximisation for sparse
  bayesian models,'' in \emph{Proc. Int. Workshop on Artificial Intelligence
  and Statistics}, Florida, USA, Jan. 2003, pp. 3--6.

\bibitem{NEEDELL2009301}
D.~Needell and J.~A. Tropp, ``{CoSaMP}: Iterative signal recovery from
  incomplete and inaccurate samples,'' \emph{Commun. ACM}, vol.~53, no.~12, pp.
  93--100, Dec. 2010.

\bibitem{matlabCoSamp}
\BIBentryALTinterwordspacing
S.~Becker, \emph{{Matlab} code: {CoSaMP} and {OMP} for sparse recovery}, Aug.
  2016. [Online]. Available:
  \url{https://de.mathworks.com/matlabcentral/fileexchange/32402-cosamp-and-omp-for-sparse-recovery}
\BIBentrySTDinterwordspacing

\bibitem{Rangan2014_AMP_EM}
U.~S. {Kamilov}, S.~{Rangan}, A.~K. {Fletcher}, and M.~{Unser}, ``Approximate
  message passing with consistent parameter estimation and applications to
  sparse learning,'' \emph{{IEEE Trans. Inf. Theory}}, vol.~60, no.~5, pp.
  2969--2985, 2014.

\bibitem{Yu_2019}
F.~{Bellili}, F.~{Sohrabi}, and W.~{Yu}, ``Generalized approximate message
  passing for massive mimo mmwave channel estimation with laplacian prior,''
  \emph{{IEEE Trans. Commun.}}, vol.~67, no.~5, pp. 3205--3219, May 2019.

\end{thebibliography}
	
	%
	%
	%
	%
	%
	
	
	

\end{document}